\documentclass[prapplied,amsmath,amssymb,superscriptaddress,floatfix,twocolumn,longbibliography]{revtex4-2}
\usepackage[final,authormarkup=none,defaultcolor=red]{changes}
\usepackage{graphicx}  
\usepackage{dcolumn}   
\usepackage{bm}        
\usepackage{placeins}
\usepackage{physics}
\usepackage{esvect}
\usepackage{upgreek}
\newcommand{\Bcal}{\bm{\mathcal{B}}}
\usepackage{url}
\usepackage[colorlinks=true,linkcolor=blue,citecolor=blue,urlcolor=blue,plainpages=false,pdfpagelabels]{hyperref}


\FloatBarrier
\begin{document}
\title{Accurate Vector Magnetometer via Zeeman Rabi Oscillations}
\author{T.~S.~Menon}
\email{thanmay.sunilmenon@colorado.edu}
\affiliation{JILA, National Institute of Standards and Technology and University of Colorado, Boulder, Colorado 80309, USA}
\affiliation{Department of Physics, University of Colorado, Boulder, Colorado 80309, USA}
\author{D.~P.~Hewatt}
\affiliation{JILA, National Institute of Standards and Technology and University of Colorado, Boulder, Colorado 80309, USA}
\affiliation{Department of Physics, University of Colorado, Boulder, Colorado 80309, USA}

\author{C.~Kiehl}
\altaffiliation{Current address: ICFO - Institut de Ci\`encies Fot\`oniques, The Barcelona Institute of Science and Technology, 08860 Castelldefels (Barcelona), Spain}
\affiliation{JILA, National Institute of Standards and Technology and University of Colorado, Boulder, Colorado 80309, USA}
\affiliation{Department of Physics, University of Colorado, Boulder, Colorado 80309, USA}

\author{M.~Ellmeier}
\affiliation{Paul M. Rady Department of Mechanical Engineering, University of Colorado, Boulder, Colorado 80309, USA}
\author{S.~Knappe}
\affiliation{Paul M. Rady Department of Mechanical Engineering, University of Colorado, Boulder, Colorado 80309, USA}
\affiliation{FieldLine Medical, Boulder CO 80301, USA}

\author{C.~A.~Regal} 
\affiliation{JILA, National Institute of Standards and Technology and University of Colorado, Boulder, Colorado 80309, USA}
\affiliation{Department of Physics, University of Colorado, Boulder, Colorado 80309, USA}

\begin{abstract}
Accurate magnetic field direction sensing in compact platforms is  critical in applications spanning magnetic navigation, space science, and biomedical imaging. We demonstrate a single-optical-axis vector optically pumped magnetometer based on  Rabi oscillations between Zeeman sublevels driven by a series of resonant radiofrequency (RF) polarization ellipses (PEs). A calibration protocol based on controlled rotations of the DC magnetic field determines the spatial orientation of each PE. We develop a detailed theoretical model describing the angular dependence of the Rabi frequencies, incorporating key systematics including RF Stark shifts and Bloch-Siegert shifts. We also account for an RF-based heading-error systematic affecting Rabi-frequency measurements arising from the nonlinear Zeeman effect. Simultaneous Larmor measurements yield the magnitude of the magnetic field, enabling integrated vector-scalar measurements. The magnetometer achieves deadzone-free vector operation with 80 $\upmu$rad mean angular accuracy and angular noise densities as low as 8 $\upmu$rad$/\sqrt{\mathrm{Hz}}$, offering a pathway towards miniaturized sensors without requiring 3D optical access or sensor rotations.
\end{abstract}

\maketitle

\section{Introduction}
 Optically pumped magnetometers (OPMs) based on spin-polarized alkali atoms are powerful tools for accurate and ultra-sensitive magnetic field measurements, achieving precision in the sub-fT regime \cite{sheng2013subfemtotesla,keder2014unshielded,dang2010ultrahigh,fabricant2023build,brown2025perspective}. Their remarkable sensitivities have enabled significant advances, ranging from tests of fundamental physics \cite{afach2021search,su2021search,ayres2021design,jackson2017constraints}, NMR spectroscopy \cite{andrews2025sensitive,bodenstedt2021fast},  unexploded ordnance detection \cite{nixon2022pulsed}, and biomedical imaging, including magnetoencephalography and magnetocardiography \cite{xia2006magnetoencephalography,borna201720,nardelli2020conformal,limes2020portable,sheng2017magnetoencephalography}. OPMs are intrinsically scalar sensors as they fundamentally operate by probing the energy spacing between magnetic sublevels, either between adjacent states within a single manifold \cite{hunter2018free,bell1961optically,schwindt2007chip} or across states belonging to different manifolds \cite{kiehl2024correcting,ben2010dead,schwindt2004chip,pollinger2018coupled}. These splittings encode only the magnitude of the DC magnetic field, leaving directional information inaccessible. Vector measurements, especially at geomagnetic field strengths, are particularly valuable in numerous applications, including magnetic navigation \cite{canciani2016absolute}, space science \cite{patton2013space,acuna2002space}, and geomagnetic surveys \cite{lu2023recent,prouty2015applications}. Extracting vector information requires coupling the atomic states to an external 3D reference frame to define the field direction accurately with respect to this reference. 

A variety of approaches have been explored to establish 3D references in vector OPMs, including multiple intersecting laser beam directions \cite{bison2018sensitive,cai2020herriott,huang2015three,afach2015highly,ingleby2018vector,patton2014all}, coil-generated magnetic fields \cite{bertrand20214he,leger2009swarm,vershovskii2006fast,pyragius2019voigt,wang2025pulsed,alexandrov2004three}, laser polarization \cite{zhang2021vector,pustelny2006nonlinear,cox2011measurements,yudin2010vector,lee1998sensitive,gonzalez2024sensitivity}, a combination of laser beam direction and a laser polarization \cite{lenci2014vectorial}, a large bias magnetic field \cite{schonau2025optically} and resonant microwave polarization ellipses \cite{kiehl2025accurate}. Each approach offers distinct advantages, ranging from enabling remote measurements through all-optical techniques to leveraging inherent electromagnetic field symmetries. However, the angular accuracy of the measurement is primarily governed by the accuracy of the mathematical model translating the atomic signals to the magnetic field orientation, and the stability of the reference system itself. In practice, references are often vulnerable to environmental drifts, beam pointing errors, and machining tolerances, limiting long-term stability and accuracy. Moreover, for many vector magnetometers, angular accuracy is not consistently benchmarked, complicating performance comparisons across platforms. These constraints are particularly critical in applications such as geophysics \cite{liu2022overview}, space science \cite{dougherty2004cassini}, and magnetic anomaly navigation \cite{mount2018navigation}, where the directional accuracy of the magnetometer is essential. Periodic recalibration of references provides a practical route to mitigating the effects of such drifts, with the recalibration interval determined by the drift rate, calibration time, and the required angular error budget.

At the same time, numerous applications also demand compact, deployable sensors. Architectures requiring multiple optical axes are especially difficult to integrate into miniaturized platforms and are particularly sensitive to axis misalignment errors \cite{pyragius2019voigt}. Single-optical-axis magnetometers are attractive as they offer a compact form factor, straightforward optical configuration, and reduce overall power requirements \cite{kiehl2025accurate,rushton2023alignment}. However, a common issue with such magnetometers is the existence of deadzones or orientations of the magnetic field where the measured signal vanishes. Probing atomic states along only one optical direction inherently leaves orthogonal spin dynamics invisible, compromising both precision and vector accuracy. Experimental approaches to overcome this issue include applying large bias fields \cite{schonau2025optically} and using machine learning algorithms to correct for orientation dependence \cite{meng2023machine}. While effective, these techniques come at the expense of added experimental complexity in the form of high power coil systems, or large training data sets leading to additional computational overhead, motivating OPM designs that offer inherently deadzone-free vector operation.

Among the various approaches for establishing vector references in OPMs, use of triaxial Helmholtz coil systems remains the simplest and most versatile \cite{lu2023recent,leger2015flight}. Here, three nominally-orthogonal coil pairs are independently driven to generate uniform magnetic fields at the vapor cell. These coil systems are also well suited to miniaturization, for instance through planar coil geometries enabled by printed circuit fabrication \cite{tayler2022miniature}. Furthermore, the coil geometry can be optically aligned to an inertial celestial reference \cite{olsen2003calibration,connerney2017juno,acuna1980magsat} directly referencing to a global stationary reference frame. Vector information is typically extracted by applying time-varying, modulated magnetic fields along the three coil axes,  often referred to as coil modulation \cite{fairweather1972vector,wang2025pulsed,alexandrov2004three}. 

A prominent example is the Helium SWARM magnetometer, which employs coil modulations to derive the DC-field vector components from their relative amplitudes in the Larmor signal, measured along a single optical axis \cite{leger2009swarm}. The simple electronic structure of $^{4}$He atoms, along with its zero nuclear spin enables highly accurate Larmor measurements, with reported scalar accuracy better than 150 pT \cite{guttin1994isotropic}. The SWARM magnetometer also reported 300 $\upmu$rad angular accuracy, which was subsequently improved to 10 $\upmu$rad after periodic calibration of coil-induced drifts through repeated sensor rotations \cite{gravrand2001calibration}. However, this approach relies on electric discharge pumping of $^4$He atoms in a 32 cm$^3$ glass-blown vapor cell \cite{rutkowski2014towards} and physical sensor rotations, which increase calibration time, hardware complexity and limit miniaturization. Additionally, amplitude-based measurements are especially vulnerable to probe-laser intensity fluctuations
arising from atomic absorption, spurious reflections, and fluctuations in the modulating field strengths. 
\begin{figure*}
    \centering
    \includegraphics[width=0.99\textwidth]{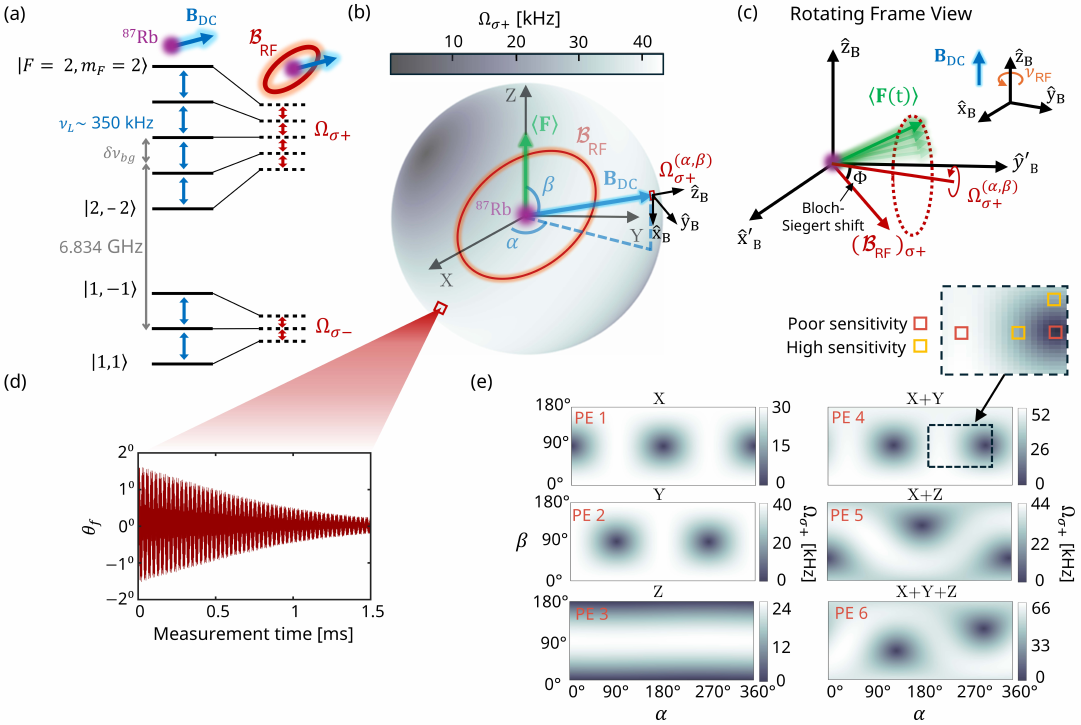}
    \caption{Vector magnetometry using Zeeman Rabi oscillations driven by resonant RF PEs. (a) Level diagram of  $^{87}$Rb in a 50 $\upmu$T DC field, showing Larmor spin precession (blue arrows) at frequency $\nu_L$. Dotted lines depict dressed states in the rotating frame, while red arrows denote Rabi oscillations under a resonant RF PE, $\Bcal_\mathrm{{RF}}$. (b) Angular dependence of $\Omega_{\sigma+}$ is shown as a function of $\mathbf{B}_{\mathrm{DC}}$ direction, $\left(\alpha,\beta\right)$ in the orthogonalized laboratory X-Y-Z coordinate system. $\langle\mathbf{F}\rangle$ denotes the initial optically pumped spin polarization. (c)  $\hat{x}'_B- \hat{y}'_B-\hat{z}_B$ denotes the rotating frame coordinate system (rotating at RF frequency, $\nu_{RF} = \omega_{RF}/2\pi$),  with $\hat{z}_B$ aligned along the DC field direction. In this frame, Rabi oscillation is visualized as spin precession driven by  $\left(\Bcal_\mathrm{RF}\right)_\mathrm{\sigma+}$. Bloch-Siegert shifts tilt the effective precession axis by $\Phi$. (d) Faraday rotation, $\theta_f$ measurement showing Zeeman Rabi oscillations. (e) Angular dependence of the Rabi frequency, $\Omega_{\sigma+}$ for the PEs used. Inset highlights a zoomed region of PE 4, showing variations in vector sensitivity determined by the angular gradient of the Rabi frequency.}
  
    \label{fig:1}
\end{figure*}

High angular accuracies have also been reported for alternative single-axis techniques utilizing electromagnetic fields as references that do not rely on physical sensor rotations. For instance laser polarization-based references using EIT resonances in glass-blown rubidium cells \cite{cox2011measurements,mckelvy2023application}  achieve $\sim\!1^\circ$ (17 mrad) level angular accuracy. More recently, vector magnetometry using microwave-driven hyperfine Rabi oscillations with a microwave polarization ellipse reference demonstrated mean angular accuracies of $\sim460\,\upmu$rad in a microfabricated rubidium cell \cite{kiehl2025accurate}.

Here, we present a single-optical-axis vector magnetometer that achieves $\sim$100 $\upmu$rad angular accuracies at geomagnetic fields, employing a coil system-based reference by leveraging Rabi oscillations in a microfabricated rubidium cell. A series of resonant RF fields drive Rabi oscillations between adjacent Zeeman sublevels in the ground-state hyperfine manifolds of $^{87}$Rb (Fig.~\ref{fig:1}a). The magnetometer exploits the dependence of the measured Rabi frequencies on the DC magnetic field to determine the direction, while additional Larmor frequency measurements, $\nu_L$ yield the magnitude. The DC magnetic field defines the quantization axis, and thereby $\sigma^\pm$, $\pi$ polarization components of the applied RF field.  Due to magnetic transition rules for the two hyperfine manifolds, $\sigma^+$  and $\sigma^-$ components of the RF field selectively excite transitions in the $F = 2$ and $F = 1$ manifolds, driving Rabi oscillations at frequencies $\Omega_{\sigma+}$ and $\Omega_{\sigma-}$, respectively  (Fig.~\ref{fig:1}a). $\Omega_{\sigma\pm}$ are determined by the projection of the polarization ellipse (PE) associated with the RF field onto the $\sigma^\pm$ polarization axes, and thus vary with the direction of the DC magnetic field (Fig.~\ref{fig:1}b). The Rabi oscillations can be further understood in the rotating frame picture (Fig.~\ref{fig:1}c), where the collective spin polarization, $\langle\mathbf{F}\rangle$ precesses at $\Omega_{\sigma+}$, driven by the effective stationary RF field component in this frame, $\left(\Bcal_\mathrm{RF}\right)_{\sigma+}$. 

Similar to the coil modulation approach described above, in our magnetometer, modulating currents are applied along the three axes of a compact triaxial coil system to produce a series of PEs.  To establish the orientation of each PE, we employ a calibration protocol based on controlled DC magnetic field rotations with a separate stable DC coil system, mitigating long-term drifts. As each measurement direction is referenced to the most recent calibration, in principle with frequent recalibration, long-term stability is fundamentally limited by the drifts in the DC coil system accumulated over the calibration time, $\mathbf{C}_t$. 
In the coil modulation approach, however, calibration via field rotations can be challenging. Here small modulation amplitudes, often chosen to limit systematics, increase the impact of DC coil technical noise directly amplified  onto the inferred direction \cite{gravrand2001calibration}. Our method relies on RF excitation resonant with the Zeeman splitting, $\nu_L$, which reduces sensitivity to low-frequency off-resonant technical noise. Moreover, by relying solely on frequency-based measurements, our technique exploits the advantage that frequencies measured against highly accurate and stable oscillators intrinsically exhibit superior signal-to-noise ratio and long-term accuracy \cite{plankensteiner2016laser}.

Angular accuracy of our magnetometer also relies on how accurately the measured Rabi frequencies are mapped to the magnetic field orientation. Achieving $\upmu$rad-level angular accuracies requires precise modeling of the RF-atom interaction at the sub-Hz-level. The RF-atom interaction is governed by a  Hamiltonian including both the DC magnetic field and the time-dependent RF field.  However, as the Rabi frequencies are appreciable relative to the RF drive frequencies (350 kHz), it is essential to go beyond the Rotating Wave Approximation (RWA). Therefore we employ a Floquet formalism \cite{shirley1965solution,florez2021floquet} to accurately model the full angular dependence of the Rabi frequencies, including key systematics such as the  Bloch-Siegert shifts that arise from counter-rotating contributions. These shifts may be viewed in the rotating frame as an additional tilt, $\Phi$ in the effective precession axis (Fig.~\ref{fig:1}c).

Additionally, we characterize a heading-error systematic affecting Rabi measurements, which we refer to as ``dynamic heading error", in analogy to a related systematic described in Ref. \cite{wang2025pulsed}. Similar to the conventional ``static" heading-error systematic \cite{bao2018suppression, lee2021heading, kiehl2024correcting} affecting Larmor frequency measurements, the underlying mechanism is the nonlinear Zeeman effect. In our case, the nonlinear Zeeman shifts lead to an additional dependence of the Rabi frequency on the field direction, and on the RF phase through the interference between Larmor and RF-driven coherences. 

We employ a series of PEs with distinct angular dependencies of the Rabi frequency (Fig.~\ref{fig:1}e), so that directions with low Rabi frequency in one PE are often compensated by high Rabi frequency in another, thereby suppressing fractional systematic errors and enhancing vector accuracy. This combined measurement also reduces variations in the Rabi frequency gradient over the full solid angle, improving overall angular resolution and ensuring deadzone-free performance despite a single axis.

We evaluate the performance of our magnetometer by comparing the vector measurements against applied magnetic fields produced by our DC coil system that is independently calibrated via Larmor spin precession measurements. Across all test magnetic field orientations, we achieved a mean angular accuracy of 80 $\upmu$rad, surpassing current  state-of-the-art  OPMs that utilize single optical axis without sensor rotation \cite{cox2011measurements,kiehl2025accurate}. We also measured a mean angular noise density of 22 $\upmu$rad/$\sqrt{\mathrm{Hz}}$, which is comparable to other high accuracy OPMs \cite{cox2011measurements,kiehl2025accurate,leger2009swarm}.   Our performance is limited by a combination of slow technical drifts in the RF system, which accumulate over the calibration and measurement sequences, as well as by residual systematics in our modeled angular dependence of the measured Rabi frequencies.

As noted earlier, the PE structure of a microwave field can also define a vector reference. This concept was previously demonstrated by our laboratory exploiting microwave-driven Rabi oscillations between hyperfine manifolds \cite{kiehl2025accurate}. The DC field direction is inferred by selectively driving $\sigma^\pm, \pi$  hyperfine transitions and extracting the corresponding Rabi frequencies. The availability of multiple, independent transitions also enables continuous tracking of slow drifts in the microwave amplitudes, and allowing self-calibration of the microwave PEs from Rabi measurements alone \cite{thiele2018self}, yielding sub-milliradian angular accuracies \cite{kiehl2025accurate}. As each hyperfine transition couples a pair of individual Zeeman sublevels, this approach also allows determination of the magnetic field magnitude without static heading error systematics \cite{thiele2018self,kiehl2024correcting}. However, this approach uses a copper microwave cavity to generate microwave PEs, leading to additional complexity and challenging miniaturization. While this work introduced simultaneous Larmor precession and Rabi driving to mitigate probing deadzones, the angular accuracy and precision degrades significantly when deviating away from the probe-beam direction. Conversely, Zeeman Rabi oscillations are driven using a  simple Helmholtz coil system that is easily compatible with existing OPM architectures and provides inherently deadzone-free vector operation. Moreover, Zeeman Rabi oscillations also experience smaller spin-exchange decoherence, resulting in coherence times approximately twice ($\sim$ 1 ms) those of hyperfine Rabi oscillations \cite{kiehl2023coherence} (Fig.~\ref{fig:1}d). A hybrid RF-microwave configuration would in principle realize an integrated vector-scalar magnetometer with minimal heading error.

\section{Experimental Setup}
\label{sec:setup}
\begin{figure}[h]
    \centering
    \includegraphics[width=0.48\textwidth]{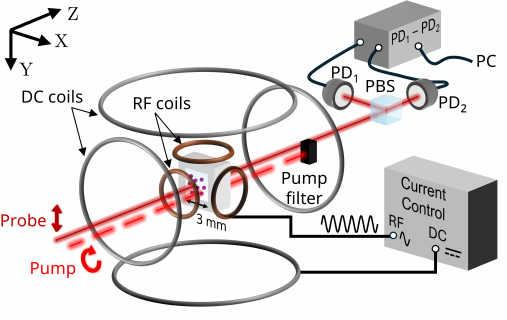}
    \caption{Experimental schematic showing co-propagating pump and probe lasers through a microfabricated $^{87}$Rb vapor cell, RF and DC coil systems, polarizing beam splitter (PBS) and photodetectors (PDs). }

    \label{fig:schematic}
\end{figure}

The experimental setup (Fig.~\ref{fig:schematic}) employs a microfabricated vapor cell ($3\times3\times2$ mm$^3$) containing $^{87}$Rb along with 350 Torr of N$_2$ buffer gas, its compact size highlighting the feasibility of sensor miniaturization. The buffer gas suppresses decoherence due to wall collisions while maintaining compatibility with Larmor measurements and  previous microwave-driven Rabi protocols \cite{kiehl2025accurate}. The cell is optically heated (Appendix \ref{sec:setup appendix}) to $100^{\circ}$C to increase vapor density. Optical heating reduces sensitivity to stray magnetic fields compared to resistive heaters. Synchronous optical pumping is performed with a 795 nm circularly polarized laser beam, amplitude modulated at the RF frequency, $\omega_{RF}/2\pi\approx\nu_L$ with 8$\%$ duty cycle, and phase-locked to the applied RF fields. The pump beam, resonant with the D$_1$ transition, is applied along the laboratory Z axis (optical axis), for 0.1 ms with a peak optical power of $\approx$ 400 mW,  polarizing atomic spins and pumping the $^{87}$Rb atoms to the stretched state $\ket{F = 2,m_F = 2}$, measured along the optical axis \cite{hewatt2025investigating,hunter2018free}. RF fields are disabled during this interval to maximize spin polarization and to suppress pump-induced systematics. Spin polarization along this axis is measured using a quantum non-demolition measurement of the Faraday rotation, $\theta_{F} \propto \langle F = 2\rangle_{Z} - \langle F = 1\rangle_{Z}$ using balanced polarization detection of a  co-propagating linearly polarized 780 nm probe beam, with an optical power of 0.8 mW and far-detuned ($\approx$ 100 GHz) from D$_2$ resonance \cite{seltzer2008developments,kiehl2023coherence} (Appendix \ref{sec:Rabi frequency extraction}). Here $\expval{F}_Z$ denotes the component of spin polarization along the laboratory Z axis (Fig.~\ref{fig:schematic}).

 A compact low-inductance triaxial Helmholtz coil system (see Appendix \ref{sec:setup appendix} for coil system details) is used to generate the resonant RF PEs that drive Rabi oscillations (Fig.~\ref{fig:schematic}).  Each coil pair is driven by an actively servoed RF current controller referenced to a function generator (Siglent 2042X) and can produce fields up to 70 $\upmu$T peak-to-peak at 350 kHz with independent amplitude and phase control. In the measurements reported here, each active RF coil pair was driven with a nominal current of $I_\mathrm{RF}\approx8$ mA$_
\mathrm{rms}$, corresponding to an estimated power of $\approx0.1$ mW (Appendix \ref{sec:setup appendix}).

For rapid, low-noise transitions between PEs, high-speed RF switches (Mini-Circuits ZASWA-2-50DRA+, $\sim$ 20 ns switching time) are employed to gate each axis individually, providing full depth amplitude modulation. The set of PEs used in this experiment is realized by selectively activating one, two or all three coils pairs simultaneously, yielding X (PE1), Y (PE2), Z (PE3), X+Y (PE4), X+Z (PE5), and X+Y+Z (PE6) configurations (Fig.~\ref{fig:1}e) in the laboratory frame. Here `+' denotes simultaneous excitation of the corresponding coil pairs. Y+Z configuration was omitted as it offered negligible additional vector information and yielded no measurable improvement in performance, while increasing the overall measurement duration.
 
 The DC magnetic field is produced by a separate Helmholtz coil system driven by an low-noise programmable current source (TwinLeaf CSB3), allowing a 50 $\upmu$T field to be applied in an arbitrary direction. The use of separate RF and DC coil systems decouples requirements for bandwidth and stability. The RF coils are compact with fewer turns to reduce inductance and enable rapid modulation for RF excitation, while a larger, higher turn-count DC coil system reduces resistive heating and improves stability and homogeneity when generating 50 $\upmu$T fields (Appendix \ref{sec:setup appendix}).  Accurate knowledge of the magnetic field is essential for the PE calibration procedure and for benchmarking vector magnetometer performance. In the latter case, magnetic field direction obtained from the measured Rabi frequencies is directly compared to the known applied field direction. To suppress background magnetic noise, the coils and the vapor cell are housed inside a four-layer $\upmu$-metal shield. Additionally, non-idealities inherent to the coil system such as imperfect coil factors (coil current-to-field conversion), axis non-orthogonalities as well as residual background fields within the $\upmu$-metal shield are accounted for through a scalar calibration of the DC coil system based solely on Larmor frequency measurements \cite{merayo2000scalar, thiele2018self,kiehl2025accurate}. The RF fields are disabled during this measurement using the high-speed RF switches to isolate the Larmor signal. The scalar calibration procedure determines the relationship between coil currents and the resulting magnetic field. Using the calibrated coil current-to-field conversion, magnetic field directions can subsequently be applied with an accuracy of 20 $\upmu$rad on timescales of at least 500 s (Appendix \ref{sec:scalar calibration}).

\section{Theoretical Model}
To accurately capture the angular dependence of the Rabi frequencies, we model the atomic spin dynamics using a density matrix formalism (see Table \ref{tab:symbols} for a summary of the symbols used throughout this work). Here, we represent the atomic state with a density matrix $\rho$ of size $(2I+1)(2S+1)$$\times$$(2I+1)(2S+1)$ where $I=3/2$ and $S = 1/2$ are the nuclear and electronic spins  of $^{87}\mathrm{Rb}$, respectively. The dynamics in the ground state can be modeled using the time-dependent Hamiltonian, $H(t)$
 \begin{multline}
H(t)=\left(A_{hfs} + h\frac{\delta\nu_{bg}}{2}\right)\mathbf{I}\cdot\mathbf{S} \,+\\
\mu_B(g_s\mathbf{S}+g_I\mathbf{I})\cdot\left(\mathbf{B}_\mathrm{DC}+\Bcal_\mathrm{RF}\mathrm{\left(t\right)}\right),
\end{multline}
 where $A_{hfs}$ is the magnetic dipole hyperfine-coupling constant, Planck's constant $h$, Bohr magneton $\mu_B$, electron and nuclear Land\'{e} g-factors $g_s$ and $g_i$,  and $\delta\nu_{bg}$ accounts for the differential frequency shift between the hyperfine manifolds induced by buffer gas collisions. The time-dependent RF field can be expressed as $\Bcal_\mathrm{RF}\mathrm{(t)} =  \mathfrak{R}\left(\bm{\mathcal{B}}_{\mathrm{0}}e^{-i\omega_{RF}t}\right)$, where $\mathfrak{R}(s)$ denotes the real part of the complex variable $s$, and the complex amplitude
 $\Bcal_0 = \sum\left(\mathcal{B}_0\right)_j\,e^{-i\phi_{j}}\,\hat{j},\;j\in \{x,y,z\}$, expressed in the laboratory frame.  We can express the total Hamiltonian as $H(t) = H_0 + H_{RF}(t)$, $H_0$ captures all the time-independent parts of the Hamiltonian. Diagonalizing $H_0$ yields the bare atomic states, which we express in the coupled angular momentum $\ket{F,m_F}$ basis with $F = 1,\,2$ in $^{87}\mathrm{Rb}$ (Fig.~\ref{fig:1}a). For a DC field $\mathrm{B_{DC}} \sim 50\;\mathrm{\upmu T}$, the energy splittings between adjacent Zeeman sublevels $\left(\ket{F,m_F}\leftrightarrow\ket{F,m_F-1}\right)$, i.e.,  the Zeeman resonance frequencies in the two manifolds, are $\sim 350\;\mathrm{kHz}$. However, the non-zero nuclear spin leads to mixing between the $\ket{F,m_F}$ states, resulting in a nonlinear Zeeman structure, where the difference between successive resonance frequencies within the manifolds differ by $\approx 36\; \mathrm{Hz}$.

The Rabi oscillations can be described by decomposing the applied RF field into its spherical components in the atomic frame. We define the spherical unit vectors $\hat{\epsilon}_{\sigma\pm} = \mp(\hat{x}_B\pm i\hat{y}_B)/\sqrt{2}$ and $\hat{\epsilon}_{\pi} = \hat{z}_B$, where $\hat{z}_B = \mathbf{B}_\mathrm{DC}/\mathrm{B}_\mathrm{DC}$ defines the quantization axis, and $\hat{x}_B, \hat{y}_B$ span the plane orthogonal to the quantization axis in the atomic frame (Fig.~\ref{fig:1}c). For a DC field direction, $\left(\alpha,\beta\right)$ in the laboratory frame, the RF amplitude, $\Bcal_0$ is projected onto the atomic frame as
\begin{equation}
\left(\mathcal{B}_0\right)^{(\alpha,\beta)}_{\sigma\pm} =\qty(R_y(-\beta) R_z(-\alpha)\,\Bcal_0)\cdot\hat{\epsilon}*_{\sigma\pm}.
\end{equation}

Here $\hat{\epsilon}*_{\sigma\pm}$ denotes the complex conjugate of $\hat{\epsilon}_{\sigma\pm}$, $R_y$ and $R_z$ are rotation matrices about Y and Z axes in the laboratory frame (Fig.~\ref{fig:1}b), and $\left(\mathcal{B}_{0}\right)^{(\alpha,\beta)}_{\sigma\pm}$ are PE components aligned with the $\sigma^\pm$ atomic polarization basis. The resulting Rabi frequencies are then approximately \cite{thiele2018self}
\begin{equation}
\Omega^{(\alpha,\beta)}_{\sigma\pm}  \approx\frac{\mu_{\sigma\pm}}{h}\left(\mathcal{B}_{0}\right)^{(\alpha,\beta)}_{\sigma\pm}.
\label{eq:Rabi}
\end{equation}
 $\mu_{\sigma\pm}$ denotes the effective magnetic transition dipole moment for the $\sigma^\pm$ transitions between adjacent Zeeman sublevels given by
\begin{align}
    \mu_{\sigma\pm}/h &= \frac{\mu_B}{2\sqrt{2}h}\langle F^\pm,m_F\pm1|g_s S_\pm+g_II_\pm|F^\pm,m_F\rangle\\&\approx5\;\mathrm{Hz/nT},\notag
\end{align}
where hyperfine manifolds are labeled $F^+ = 2$ and $F^- = 1$, and $S_\pm = S_x\pm iS_y$, $I_\pm = I_x\pm iI_y$ are defined in the atomic frame, $\{x_B,y_B,z_B\}$ (Fig.~\ref{fig:1}b). $\mu_{\sigma\pm}$ is related to the gyromagnetic ratio, $\gamma$ by $\mu_{\sigma\pm}/h = \gamma/\sqrt{2}$, where the factor of $\sqrt{2}$ follows from the normalization convention of $\sigma^\pm$ components.

Optical pumping into the $F = 2$ manifold yields substantially higher signal-to-noise ratios for the $\sigma^+$-driven Rabi signals.  Although population transfer to $F = 1$ via a microwave adiabatic rapid passage pulse would enable $\Omega_{\sigma -}$ measurements, our proof-of-principle implementation here does not rely on microwave control, and we restrict Rabi measurements to $\Omega_{\sigma +}$. In subsequent sections, we employ a theoretical model to capture the angular dependence of the Rabi frequencies,  $\Omega_{\sigma+}\left(\Bcal_0,\alpha,\beta\right)$ on the DC field direction. We first calibrate the six PEs by measuring their Rabi frequencies at various known magnetic field directions and fit the theoretical model to the measurements to extract the corresponding PE parameters. We then use Rabi measurements driven by the calibrated PEs, together with the same theoretical model to infer an unknown magnetic field direction. However, the minimal model described in Eq.~\eqref{eq:Rabi} is insufficient to achieve high angular accuracy as it neglects several systematic effects discussed below.

 \subsection{Modeling Bloch-Siegert and RF Stark Shifts}
\begin{figure}[thb]
    \centering
    \includegraphics[width=0.45\textwidth]{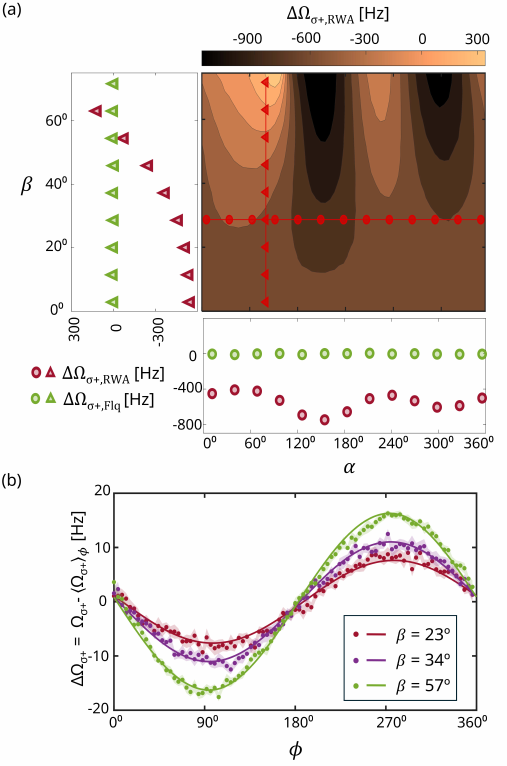}
    \caption{Systematic effects in Rabi frequency measurements and comparison with theory. (a) Residuals between measured Rabi frequency and predictions from a RWA model, $\Delta\Omega_{\sigma+,\mathrm{RWA}}$ as a function of magnetic field direction. Bottom and left panels show residuals from the RWA model and a Floquet-based model, $\Delta\Omega_{\sigma+,\mathrm{Flq}}$  along the horizontal and vertical cuts at fixed $\beta$ = 28.65$^\circ$ and $\alpha = 80.21^\circ$, respectively. Measurement uncertainties are smaller than the markers. (b) Variation of $\Omega_{\sigma+}$ with the applied RF phase, $\phi$ for three polar angles. $\alpha=0$ for all three directions. Results of the model (solid lines) accounting for dynamic heading errors in the PE, $\Bcal_\mathrm{RF}\sim 9.1 \cos(\omega_{RF}t+\phi_0+\phi) \hat{x}\; \upmu \mathrm{T}$ with technical offset phase, $\phi_0 \sim 232^\circ$.}

    \label{fig:systematics}
\end{figure}

Determining the Rabi frequency conventionally involves transforming $H(t)$ into a frame rotating at the RF frequency, $\omega_{RF}/2\pi = 350$ kHz (Fig.~\ref{fig:1}c).  Applying the Rotating Wave Approximation (RWA) then eliminates counter-rotating terms oscillating at 2$\omega_{RF}$, resulting in a simplified time-independent Hamiltonian. In this rotating frame, co-rotating but off-resonant couplings to neighboring Zeeman transitions produce RF Stark shifts, which can be treated within the RWA. However, the same secord-order AC Stark mechanism also produces shifts from the counter-rotating terms, conventionally referred to as Bloch-Siegert shifts \cite{bloch1940magnetic}. Because the RWA discards the coupling to counter-rotating fields, this contribution is neglected. These shifts also scale quadratically with the RF field amplitude and lead to measurable deviations in the Rabi frequency \cite{abel2020optically,beato2020second,sudyka2019limitations}. As $ \Omega_{\sigma\pm}/(\omega_{RF}/2\pi)$ can be as large as $\simeq 0.2$ in our system, accurate modeling of Rabi dynamics requires these effects to be explicitly accounted for in our model \cite{florez2021floquet}. Fig.~\ref{fig:systematics}a shows the residuals between $\sigma^+$ Rabi frequency measurements and predictions from a time-independent Hamiltonian model under the RWA, $\Delta\mathrm{\Omega_{\sigma+,RWA}}$ as a function of magnetic field orientation. Measurements were performed using a PE characterized by the following amplitudes and phases in the laboratory frame: $\{(10.062\;\upmu \mathrm{T},-46^\circ)_x,\allowbreak (12.716\;\upmu \mathrm{T},137^\circ)_y,\allowbreak (7.889\;\upmu \mathrm{T},-155^\circ)_z\}$. The residuals, on the order of 100s of Hz, arise primarily from Bloch-Siegert systematics unaccounted for in the RWA.

We therefore employ a Floquet formalism \cite{shirley1965solution, sun2022multi,shi2024high} that expands the Hilbert space to encompass both the bare atomic states and a ladder of harmonic states associated with the RF field. This formalism captures the full periodic structure of the Hamiltonian and incorporates both co-rotating RF Stark shifts and the counter-rotating Bloch-Siegert contributions within a single dressed-state description. In Fig.~\ref{fig:systematics}a, the horizontal and vertical slices compares the residuals $\mathrm{\Delta\Omega_{\sigma+,RWA}}$ and $\mathrm{\Delta\Omega_{\sigma+,Flq}}$, where the latter denotes the difference between Rabi frequency measurements and predictions from the Hamiltonian exploiting the Floquet framework. The Floquet Hamiltonian is able to effectively capture these systematics, yielding residual discrepancies between the predicted and the measured Rabi frequencies, $\Delta\Omega_\mathrm{\sigma+,Flq}$  to $\lesssim3.5$ Hz in Fig.~\ref{fig:systematics}a. 

As the RF field drives transitions within the $F = 1$ and $F=2$ manifolds independently, we can partition the Hamiltonian as $H(t) = H_{F=1}(t) + H_{F=2}(t)$, and solve for the dynamics within each manifold separately. Within the Floquet framework, the time-dependent periodic Schrödinger equation,
\begin{equation}
\label{eq:sch}
i\hbar\frac{\partial\psi_F(t)}{\partial t} = H_F(t)\psi_F(t),
\end{equation}
has solutions of the form
\begin{equation}
\label{eq:psi}
\psi_F(t) = e^{-iqt/\hbar}\ket{\chi(t)},
\end{equation}
where $\ket{\chi(t)}$ is periodic in time and $q$ is defined as the associated quasienergy of $\ket{\chi(t)}$.

In the Floquet treatment, we transform the periodic time-dependent Hamiltonian, $H_F(t)$ into an equivalent time-independent infinite-dimensional Floquet Hamiltonian, $\widetilde{H}_F$ \cite{shirley1965solution}. The eigenvalues of $\widetilde{H}_F$  describe the quasienergy spectrum, $q$ with the corresponding eigenstates, $\ket{\chi(t)}$. To construct $\widetilde{H}_F$, we first decompose the time-dependence of the Hamiltonian as well as the quasienergy eigenfunctions into various harmonics of $\omega_{RF}$. For a periodically driven Hamiltonian with period, $2\pi/\omega_{RF}$, $H_F(t)$  and $\ket{\chi(t)}$ take the form \cite{son2009floquet}
\begin{align}
H_F(t)&=\sum_{n=-\infty}^{+\infty}H_F^{(n)}e^{in\omega_{RF}t},\notag\\\ket{\chi(t)}&=\sum_{n=-\infty}^{+\infty}\ket{\chi^{(n)}}e^{in\omega_{RF}t}.
\label{eq:H}
\end{align}

Because our applied RF field is purely sinusoidal, only the static and first-harmonic components contribute, such that $H_F^{(n)}=0$ for $\left|n\right|>1$. Following Floquet-state nomenclature convention, we express $\widetilde{H}_F$ in the basis
\begin{equation}
\ket{F,m_F;n} = \ket{F,m_F}\otimes\ket{n},
\end{equation}
where $n$, spanning from $-\infty$ to $+\infty$, denotes the Floquet harmonic index corresponding to the $n^{th}$ Fourier component at frequency $\mathrm{\omega_{RF}}$.

Substituting Eqs.~(\ref{eq:psi}) and (\ref{eq:H}) in  Eq.~\eqref{eq:sch}, we obtain the following  equation for the matrix elements of $\widetilde{H}_F$
\begin{multline}
\langle F,m_F';n'|\widetilde{H}_F|F,m_F;n\rangle = \langle F,m_F'|H^{(n'-n)}_F |F,m_F\rangle \\
+ n\hbar\omega\,\delta_{m_F',m_F}\delta_{n',n}.
\label{eq:Flqmatrix}
\end{multline}

Here the first term corresponds to off-diagonal couplings resulting from RF-driven transitions between adjacent Zeeman sublevels, while the diagonal term introduces an energy shift proportional to the harmonic index, n. Eq.~\eqref{eq:Flqmatrix} highlights the distinction between the Floquet and rotating wave pictures. In the RWA, only the resonant Fourier component of the driving field is retained while the other is neglected. In the Floquet formalism, all spectral components are preserved, thereby capturing both the resonant transitions as well as the off-resonant effects, such as the Bloch-Siegert shifts. The quasienergy spectrum $q_{F,m_F;n}$ is obtained by solving the eigenvalue equation for $\widetilde{H}_F$
\begin{align}
\sum_{m_F''} \sum_{n''} 
    \langle F, m_F'; n'|& \widetilde{H}_F |F, m_F''; n'' \rangle 
    \langle F, m_F''; n''|q_{F,m_F;n} \rangle\notag \\
    &= q_{F,m_F;n} \langle F,m_F';n'|q_{F,m_F;n}\rangle,
\end{align}
with eigenstate $\langle F,m_F';n'|q_{F,m_F;n}\rangle=\braket{F,m'_F}{\chi^{(n'-n)}_{F,m_F}}$.

Eigenvalues of $\widetilde{H}_F$ can be numerically solved by truncating the number of Floquet blocks. For $\Omega_{\sigma\pm}/(\omega_{RF}/2\pi) \simeq 0.2$, we have found that $n =-5,\ldots,+5$ is sufficient for capturing the Bloch-Siegert systematics relevant to our measurements. As the dressed-state solutions of the Hamiltonian, $H_F(t)$ map directly onto the eigenstates of the Floquet Hamiltonian, $\widetilde{H}_F$, the Rabi frequencies follow directly from the quasienergy splittings \cite{shirley1965solution}. The Rabi frequency $\Omega_{\sigma\pm,m}$ for the transition $\ket{F^\pm,m\pm1}\leftrightarrow\ket{F^\pm,m}$ is then given by

\begin{equation}
\Omega_{\sigma\pm,m}= \frac{q_{F^{\pm},m;-m} - q_{F^{\pm},(m-1);-(m-1)}}{h}.
\label{eq:eig}
\end{equation}

The Floquet spectrum therefore determines the full set of Zeeman Rabi frequencies in the driven system. However, deviations from linear Zeeman spacing introduce additional systematics which must be included to accurately model the measured Rabi frequencies.

\subsection{Modeling Dynamic Heading Error Systematics from Nonlinear Zeeman Effect}

 Within the $F = 2$ manifold, due to the unequal energy level spacings introduced by the nonlinear Zeeman effect, the four $\Delta m_F = 1$ transitions exhibit slightly differing $\Omega_{\sigma+,m}$. The splittings between adjacent $\Omega_{\sigma+,m}$, on the order of tens of Hz, depend on the spatial structure of PE as well as the DC magnetic field strength.  As the decoherence rate, $\Gamma\sim2\pi\times1$ kHz of the measured Rabi signal far exceeds the splitting between the Rabi frequencies, the individual components remain unresolved. The effective measured Rabi frequency  $\Omega_{\sigma+}$ therefore arises as a weighted average over the $\Omega_{\sigma+,m}$. This situation closely resembles the ``static" heading error systematic in scalar OPMs  \cite{lee2021heading,kiehl2024correcting,bao2018suppression}. Unequal and unresolved Zeeman splittings cause the  Larmor frequency, $\nu_L$ to be a superposition over the individual Zeeman frequencies. Therefore, the measured Larmor frequency depends on the projection of the spin, $\langle\mathbf{F}\rangle$ onto $\mathbf{B}_{\mathrm{DC}}$, producing a direction-dependent error in the resulting DC field magnitude measurement. Similarly, relative variations in the contribution of each $\Omega_{\sigma+,m}$ distort the observed Rabi lineshape, giving rise to the dynamic heading error \cite{wang2025pulsed}.  This systematic can be understood by transforming to a frame co-rotating with $\left(\Bcal_\mathrm{RF}\right)_{\sigma+}$. As illustrated in Fig.~\ref{fig:1}c, in this frame, $\left(\Bcal_\mathrm{RF}\right)_{\sigma+}$ appears stationary and the atomic spin undergoes coherent precession at the Rabi frequency. As a result, similar to the static heading error, the effective measured Rabi frequency, $\Omega_{\sigma+}$ becomes sensitive to the projection of atomic spin onto $\left(\Bcal_\mathrm{RF}\right)_{\sigma+}$.
   
 We can characterize this effect experimentally by varying the phase of the driving RF field relative to the synchronous optical pumping modulation pulses, which are phase-locked to the RF oscillator. Varying this phase changes the relative timing between the RF waveform and the pump modulation pulses. In the rotating frame, this phase shift corresponds to a rotation of the stationary RF component, $\left(\Bcal_\mathrm{RF}\right)_{\sigma+}$, relative to the pumped spin polarization (Appendix \ref{sec:RF_phase}). As a result, the effective spin projection onto
 $\left(\Bcal_\mathrm{RF}\right)_{\sigma+}$ varies, thereby inducing small but measurable shifts in the Rabi frequency. Fig.~\ref{fig:systematics}b shows the variation of  $\Omega_{\sigma+}$ with the applied RF phase, $\phi$ for $\Bcal_\mathrm{RF}\sim9.1 \cos(\omega_{RF}t+\phi + \phi_0) \,\upmu \mathrm{T}\; \hat{x}$, revealing a sinusoidal variation characteristic of heading error. Here $\phi_0$ denotes technical phase shifts arising due to the RF current controller and the coil inductance, while $t=0$ marks the transition from optical pumping to RF excitation. $\phi_0 +\phi$ defines the relative phase between the pump modulation and the driving field. Mathematically, $\phi+\phi_0 = \omega_{RF}\Delta t$, where $\Delta t$ denotes the time delay between the rising edge of the pump modulation and peak of the RF waveform. 

 We model this systematic by calculating the weighted contributions of the individual Rabi frequencies, $\Omega_{\sigma+,m}$ for a given RF phase based on the initial optically pumped state. The atomic dynamics during optical pumping can be simulated using a standard optical pumping model (Appendix \ref{sec:pump}). We model the process using the quantum master equation including a pumping term, $R_{OP}\left(\varphi_{n}\left(1+2S_z\right)-\rho\right)$, where $R_{OP}$ is the optical pumping rate determined on the pump power and detuning from atomic resonance,  and  $\varphi_n = \rho/4+\mathbf{S}\cdot\rho\mathbf{S}$ denotes the nuclear component of the density matrix, $\rho$ \cite{seltzer2008developments}. If $\rho_{0}$ is the resulting density matrix after the pumping sequence, we can simulate the Faraday rotation signal, $\theta_f(t)$ 
 \begin{equation}
    \theta_f(t) = \mathrm{Tr}\left(e^{-i\widetilde{H}_{F}t/\hbar}\rho_0e^{i\widetilde{H}_{F}t/\hbar}\Theta_f\right),
\end{equation}
   where $\Theta_f$ is the Faraday rotation operator \cite{kiehl2023coherence} and $\mathrm{Tr(..)}$ denotes the trace operation. Expressed in the Floquet eigenbasis, $\{|q_{F,m_F;m}\rangle\}$, this equation yields a sum over coherences oscillating at frequencies given by differences of the Floquet quasienergies,
   \begin{align}
         \mkern-9mu\theta_f(t) = \mkern-17mu\sum_{m_F,m'_F,m',m}\mkern-30mu&\mel{q_{F,m_F;m}}{\rho_{0}}
         {q_{F,m'_F;m'}}\notag\\
         &\times\mel{q_{F,m'_F;m'}}{\Theta_f}{q_{F,m_F;m}}\notag\\
         &\times \exp\left(-i\left(q_{F,m'_F;m'} - q_{F,m_F;m}\right)t/\hbar\right).
  \end{align}
 
 The Fourier components of $\theta_f(t)$ at the frequency $\Omega_{\sigma+,m}$ have the amplitude, $A_m$ given by magnitudes of the corresponding matrix elements (Eq.~\eqref{eq:eig}),
\begin{equation}
A_m =\left|
\begin{aligned}
&\mel{q_{F^+,m;-m}}{\rho_{0}}{q_{F^+,(m-1);-(m-1)}}\\&\hspace{1em}\times\mel{q_{F^+,m;-m}}{\Theta_f}{q_{F^+,(m-1);-(m-1)}}
\end{aligned}
\right|.
\label{eq:Amplitudes}
\end{equation}

 The effective Rabi frequency, $\Omega_{\sigma+}$ is then modeled as a weighted average over $\Omega_{\sigma+,m}$

 \begin{equation}
    \Omega_{\sigma+} = \frac{\sum_{m} A_{m}\Omega_{\sigma+,m}}{\sum_m A_m},
    \label{eq:dynamic_he}
 \end{equation}
where $m\in\{-1,0,1,2\}$, labels the four Zeeman resonances in $F=2$. Solid lines in Fig.~\ref{fig:systematics}b show that Eq.~\eqref{eq:dynamic_he} captures the modulation of $\Omega_{\sigma+}$ with RF phase in excellent agreement with measurements to within $\sim\,$3 Hz using a pumping rate, $R_{OP} = 0.7$ MHz. The reduction in $\Omega_{\sigma+}$ variation for smaller $\beta$ mirrors the reduction in static heading error as $\mathbf{B}_\mathrm{DC}$ becomes orthogonal to the pumped spin polarization. In the Rabi case, as the DC magnetic field approaches the pump beam axis, the pumped  spin polarization becomes increasingly orthogonal to the $\sigma^+$ polarization axis. In this limit, phase-dependent variation of the spin projection in the rotating frame is reduced, leading to a smaller dynamic heading error. 

\begin{figure*}
    \centering
    \includegraphics[width=0.95\textwidth]{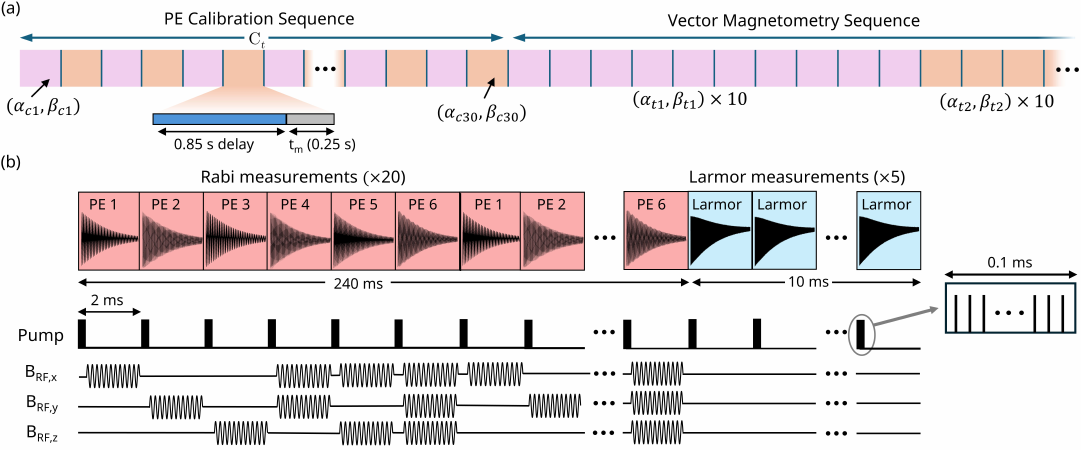}
    \caption{Schematic of the measurement protocol illustrating PE calibration and vector magnetometry sequences. (a) Timing diagram illustrating the sequence of the applied 50 $\upmu$T DC magnetic field orientations. During calibration, Rabi and Larmor frequencies are measured at thirty random but predefined magnetic field directions, $\left(\alpha_{cj},\beta_{cj}\right)$, followed by measurements at random test directions, $\left(\alpha_{tj},\beta_{tj}\right)$, each repeated 10 times.  Each Rabi-Larmor measurement cycle, including the 10 repeated measurements at every test direction, is preceded by a 0.85 s delay to allow the DC currents to settle while maintaining uniform timing across repeats. (b) For each DC field orientation, atoms are interrogated using six PEs generated by distinct permutations of RF pulses applied to the RF coil system.  $\mathrm{B}_{\mathrm{RF,i}}, i\in \{x,y,z\}$ denotes the RF signals applied to the corresponding coil pair. To reduce technical noise, each Rabi measurement is repeated and averaged 20 times, while each Larmor measurement is averaged over 5 repetitions.}  
    \label{fig:timing}
\end{figure*}

\section{PE Calibration}
\label{sec:calibration}

Calibration of all six PEs constitutes the essential first step towards mapping the observed Rabi frequencies to the underlying magnetic field direction. PE calibration is performed by measuring $\Omega_{\sigma +}$ for all six PEs for thirty random but predefined magnetic field directions.
Sampling substantially more directions than the six PE fit parameters helps ensure that the fitted parameters reflect the global angular structure of the PE, instead of matching a limited set of measurements. This approach also reduces the tendency for residual modeling errors to be absorbed into the fitted PE parameters. This choice represents a practical balance for the setup. Using additional directions increase the calibration time, $\mathbf{C}_t$ and therefore susceptibility to slow RF drifts, while fewer directions reduce the angular coverage and increase the sensitivity of the fit to systematic bias. Our experiment protocol (Fig.~\ref{fig:timing}) consists of a PE calibration sequence followed by the vector magnetometry sequence. The coherence times of the measured Faraday rotation signals (Fig.~\ref{fig:1}d) are limited to $\sim$ 1 ms, primarily due to relaxation from spin-exchange and spin-destruction collisions between $^{87}$Rb atoms, and wall collisions in the microfabricated cell.  
 
Each measurement cycle begins with 0.1 ms of synchronous optical pumping, followed by a 1.9 ms RF pulse across one or more coil pairs to drive Rabi oscillations. For Larmor measurements, the RF fields remain switched off for the same duration. Rabi and Larmor frequencies are extracted by least-squares fitting  of the Faraday rotation signals (Appendix \ref{sec:Rabi frequency extraction}). For each magnetic field orientation, Rabi measurements are repeated 20 times per PE, while Larmor measurements are repeated 5 times to suppress technical noise, yielding a total measurement time, $t_m$ of 250 ms. 

The Rabi measurements are then fit to our theoretical model (Eqs.~\eqref{eq:eig} and \eqref{eq:dynamic_he}) introduced in the previous section. The parameters of each PE are obtained by minimizing the cost function
\begin{equation}
\mathcal{C}(\Bcal_0) = \sum_{j=1}^{30}\left[\widetilde{\Omega}_{\sigma+}\left(\Bcal_0,(\alpha,\beta)_j\right) - \left(\Omega_{\sigma+}\right)_j\right]^2.
\end{equation}

Here $\widetilde{\Omega}_{\sigma+}\left(\Bcal_0,(\alpha,\beta)_j\right)$ denotes the Rabi frequency prediction from our theoretical model  evaluated at each magnetic field orientation. $\Bcal_0$ is comprised of the three amplitudes and phases, $\{(\mathcal{B}_\mathrm{0})_i,\phi_i\}, i\in\{x,y,z\}$ in the laboratory frame. The calibrated field, $\Bcal_\mathrm{0,cal}$ is determined by applying a nonlinear least squares optimization algorithm (Levenberg-Marquadt) to minimize $\mathcal{C}(\Bcal_0)$. The complete  calibration sequence for all six PEs currently requires a time,  $\mathbf{C}_t\sim$\, 30 s, primarily constrained by the 0.85 s applied delay,  which allows DC currents to settle to within 10 ppm of the current setpoint. With faster current control, this dead time could be reduced to $\sim$ 0.05 s, allowing $\mathbf{C}_t$ to be reduced to 9 s. 

The calibration results in root mean square residuals, $\sqrt{\mathcal{C}(\Bcal_\mathrm{0,cal})}$, ranging from 1 Hz (0.2 nT) to  3.5 Hz (0.7 nT) across the six PEs. These residuals quantify the discrepancy between the calibrated PE model and the measured Rabi frequency at the calibration directions, and therefore provide a measure of calibration error that propagates into the subsequent vector reconstruction.  They arise in part from slow drifts in the RF system during the calibration procedure (Appendix \ref{sec:Drift}). Although the DC magnetic field system remains stable over repeated measurements, the RF coil system and the associated electronics drift over the calibration time, which is extended in large part by the measurement delays applied to allow DC fields to settle after each orientation change (Fig.~\ref{fig:drift}).  Additional deviations up to a few Hz, arise from systematics not included in the model presented here and therefore contribute to the observed calibration residuals. In particular, these include systematics due to spin-exchange collisions and residual population in the $F = 1$ manifold. These systematics couple to the $\sigma^-$ component of the applied PEs and shift the effective observed Rabi frequency (Appendix \ref{sec:exact}).

\section{Vector Magnetometry Evaluation}
\label{sec:eval}

\begin{figure*}[t]
    \centering
    \includegraphics[width=0.92\textwidth]{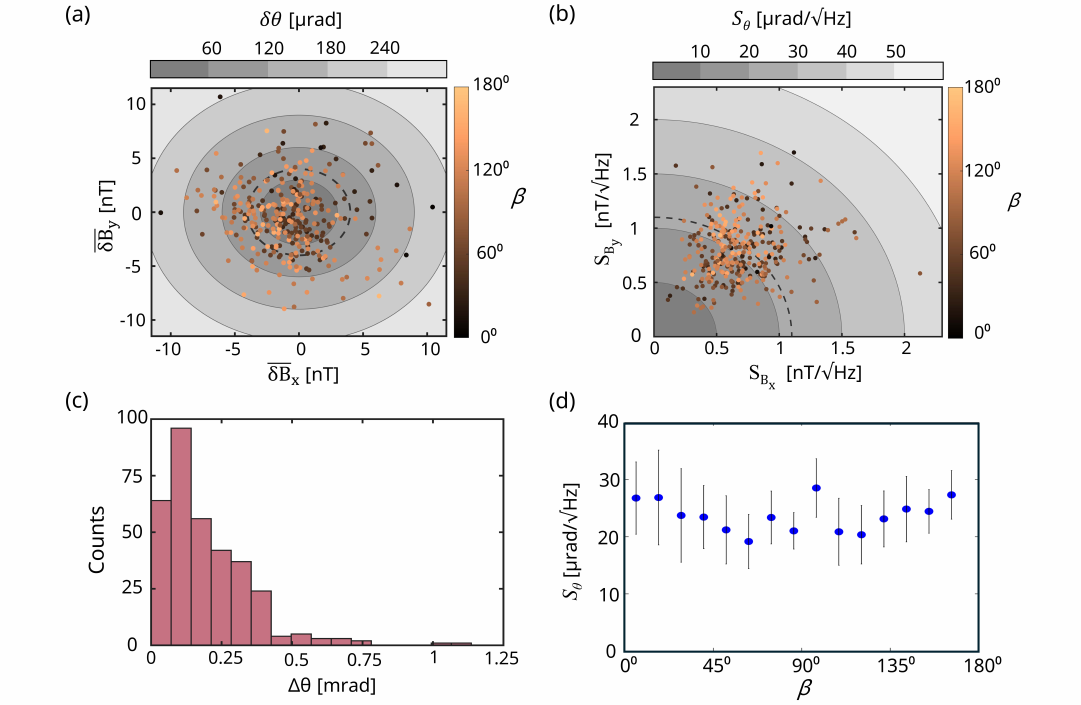}
    \caption{Evaluation of vector magnetometer performance. (a) Transverse component errors, $\overline{\updelta\mathrm{B}}_\mathrm{x}$, $ \overline{\updelta\mathrm{B}}_\mathrm{y}$  and (b) transverse component noise densities $\mathrm{S_{B_x},S_{B_y}}$  for more than 340 randomly sampled magnetic field orientations distributed over the unit sphere. Each point corresponds to a distinct applied field direction, with non-uniform angular spacing, and is color coded by polar angle, $\beta$. Gray contours indicate boundaries of constant angular accuracy, $\delta\theta$ and angular noise density, $S_{\theta}$, and dashed lines mark the mean angular accuracy (80 $\upmu$rad) and noise density (22 $\upmu$rad$/\sqrt{\mathrm{Hz}}$) over the test directions in (a) and (b), respectively. (c) Histogram shows the relative angular error, $\mathrm{\Delta\uptheta}$ between vector measurements obtained using disjoint sets of PEs. (d) Angular noise density, $S_{\theta}$ binned as a function of the polar angle, $\beta$, illustrating the variation of noise with magnetic field orientation.}

    \label{fig:data}
\end{figure*}

Once calibrated, the six PEs are utilized to determine an unknown magnetic field direction, while the field magnitude is obtained independently from Larmor measurements. For a given field direction, the calibrated PEs drive Zeeman Rabi oscillations, and the corresponding Rabi frequencies are extracted from the measured Faraday rotation signals. The unknown magnetic field direction is obtained by minimizing the cost function \cite{kiehl2025accurate}
\begin{equation}
\mathcal{C}(\alpha,\beta) = \sum_{i=1}^{6}\left[\widetilde{\Omega}_{\sigma+}\left(\left(\Bcal_\mathrm{0,cal}\right)_i,\, \alpha,\beta\right) - \left(\Omega_{\sigma+}\right)_i\right]^2,
\label{eq: cost direction function}
\end{equation}
where $\widetilde{\Omega}_{\sigma+} (...)$ denotes the Rabi frequency predicted from the theoretical model in Eqs.~\eqref{eq:eig} and \eqref{eq:dynamic_he}. The set $\left(\Bcal_\mathrm{0,cal}\right)_i, \,i\in \{1,2,...,6\}$ corresponds to the six calibrated PEs. Similar to the calibration procedure, minimization of the cost function, $\mathcal{C}(\alpha,\beta)$  via a least squares algorithm yields the magnetic field direction $(\alpha^*,\beta^*)$.  Although the present implementation uses only $\Omega_{\sigma+}$ measurements, employing six distinct PEs imposes multiple independent constraints on the two unknown angles, yielding a unique solution for the direction and enhancing overall angular precision.

We evaluate the angular accuracy (Fig.~\ref{fig:data}a) and noise density (Fig.~\ref{fig:data}b) of the vector magnetometer by comparing the extracted field direction, $(\alpha^*,\beta^*)$ to the known applied magnetic field orientation \cite{mckelvy2023application,kiehl2025accurate,amtmann2024accuracy}.  To characterize performance, 50 $\upmu$T DC magnetic fields were applied along 342 directions randomly distributed over the full solid angle. These directions provide a comprehensive angular coverage for evaluating vector reconstruction across different field orientations. Recalibration measurements were interleaved nearly once every 50 test directions. This interval was chosen to reduce the accumulation of slow RF drifts, while limiting the overhead associated with repeated PE calibration (Appendix \ref{sec:Drift}). For each direction, Rabi frequencies corresponding to the six PEs were measured and the direction, $(\alpha^*,\beta^*)$ extracted using the procedure described above.  Furthermore, each Rabi-Larmor measurement cycle was repeated 10 times per direction (Fig.~\ref{fig:timing}a), with the same 0.85 s delay applied before each cycle.

To facilitate comparison with other vector magnetometer platforms, we express our angular accuracy in terms of transverse component errors, $\left(\updelta\mathrm{B}_x,\updelta\mathrm{B}_y\right)$ relative to the applied field direction. To uniformly characterize performance for all field directions and prevent issues with azimuthal angle, $\alpha$ becoming ill-defined as $\beta\rightarrow0^{\circ}$ or $180^{\circ}$, we rotate the measured magnetic field towards the z-axis, following the approach introduced in Ref.~\cite{kiehl2025accurate},
\begin{equation}
    \mathrm{\left(\updelta B_x,\updelta B_y,B_z\right)} = R_y(-\beta) R_z(-\alpha) \mathbf{B}_\mathrm{m}.
\end{equation}

Here $ \mathbf{B}_\mathrm{m}$ denotes the measured magnetic field vector oriented along the extracted direction $\left(\alpha^*,\beta^*\right)$, while $\left(\alpha,\beta\right)$ is the applied magnetic field direction obtained from the scalar coil calibration described in Sec.~\ref{sec:setup}. The angular accuracy, $\delta\theta$ given by
\begin{equation}
\delta\theta = \tan^{-1}\left(\sqrt{\overline{\updelta\mathrm{B}}^2_\mathrm{x} + \overline{\updelta\mathrm{B}}^2_\mathrm{y}}/\mathrm{B}_\mathrm{m}\right),
\end{equation}
where  $\overline{\updelta\mathrm{B}}_\mathrm{x}$ and $\overline{\updelta\mathrm{B}}_\mathrm{y}$, denote the average transverse component errors, averaged over 10 repetitions for each direction and $\mathrm{B_m} = |\mathbf{B}_\mathrm{m}|$. Averaged over all directions, the magnetometer yields a mean angular accuracy of $\langle\delta\theta\rangle = 80\,\upmu$rad, which corresponds to a transverse field accuracy of 4 nT (Fig.~\ref{fig:data}a). The observed accuracy is consistent with the magnitude of the calibration residuals obtained for the six PEs. 

While the analysis above quantifies the vector accuracy with respect to the calibrated DC coil system, we can also perform an internal consistency check on the vector magnetometer by comparing measurements obtained from two disjoint sets of PEs. Using a smaller PE subset generally degrades the overall angular precision and accuracy, as the measurements become more vulnerable to angular regions with weak Rabi frequency gradients and to systematics associated with smaller Rabi frequencies. However, this approach enables an evaluation of vector performance that does not rely on an external scalar calibration for DC coil system. The relative angular error, $\Delta\uptheta$ defined below
\begin{equation}
\Delta\uptheta = \cos^{-1}{\left(\frac{\overline{\mathbf{B}}_\mathrm{m1}\cdot\overline{\mathbf{B}}_\mathrm{m2}}{\mathrm{B}^2_\mathrm{m}}\right)}.
\end{equation}

Here $\overline{\mathbf{B}}_\mathrm{m1}$  and $\overline{\mathbf{B}}_\mathrm{m2}$ are the measured magnetic field vectors averaged over 10 repetitions utilizing the PE subsets $\{1,2,6\}$ and $\{3,4,5\}$, respectively. These subsets were chosen to have comparable average Rabi frequencies across all the measured directions, ensuring a fair comparison. $\Delta\uptheta$ was measured for the same 342 directions measured in Fig.~\ref{fig:data}a and the resulting distribution is shown in Fig.~\ref{fig:data}c. The mean relative angular error, $\langle\Delta\uptheta\rangle$, measured across all directions was 197 $\upmu$rad. This discrepancy arises primarily from systematics not fully captured by the Floquet model, which impact the two PE subsets differently for a given direction. The effects of these systematics are significantly more pronounced due to the smaller size of the PE subsets. Additional contributions arise from relative amplitude and phase drifts between the PE subsets, particularly for PEs composed of superpositions of RF fields from multiple coil axes.

The angular noise density of the magnetometer is calculated from the fluctuations of the measured transverse component errors over the 10 repetitions at each magnetic field orientation taking the bandwidth into account. Assuming white noise, the component noise density, $\mathrm{S_{B_j}}$ is given by
\begin{equation}
    \mathrm{S_{B_j}} = \sigma_{B_j}\sqrt{2t_{m}}\,,
\end{equation}
where $\sigma_{B_j}$ is the standard deviation of $\updelta\mathrm{B}_\mathrm{j}$  across the 10 repeated measurements, and $t_{m}$ denotes our fundamental measurement time of 0.25 s. Because each vector estimate is obtained from fitting the Zeeman Rabi frequencies to the Faraday rotation signals (Appendix \ref{sec:Rabi frequency extraction}), the measured angular noise density naturally incorporates  fluctuations arising from the finite coherence time as well as the orientation-dependent signal-to-noise ratio of the six PEs for each test direction. This calculation excludes the additional 0.85 s technical delay before measurement, shown in Fig.~\ref{fig:timing}a. The angular noise density is then given by, $S_\theta = \tan^{-1}\left(\sqrt{\mathrm{S^2_{B_x}+S^2_{B_y}}}/\mathrm{B_m}\right)$ \cite{kiehl2025accurate}. Because the Rabi-frequency sensitivity associated with the six PEs depend on the orientation of the applied field (Fig.~\ref{fig:1}e), the angular noise density is inherently  direction dependent. As a result, $S_\theta$ varies across the whole unit sphere (Fig.~\ref{fig:data}b), reaching a minimum of 8 $\upmu$rad$/\sqrt{\mathrm{Hz}}$ at $(\alpha,\beta) = (65.57^{\circ},49.68^{\circ})$.  Averaging over all measured directions, we obtain a mean angular noise density $\langle S_\theta\rangle$ is  22 $\upmu$rad$/\sqrt{\mathrm{Hz}}$, equivalent to a transverse component noise density of $1.1\,\mathrm{nT}/\sqrt{\mathrm{Hz}}$. 

The measured noise density is comparable to that of previously reported single-axis vector magnetometers operating at geomagnetic fields \cite{gravrand2001calibration,kiehl2025accurate}, though alternative architectures employing multi-axis configurations \cite{wang2025pulsed}  or approaches that rely on low-noise active field compensation coils to realize an effective zero-field environment at the vapor cell \cite{bertrand20214he} have demonstrated lower noise floors. Our noise density is primarily limited by shot-to-shot residual amplitude variations in the applied RF fields, arising from short-term thermal drifts as well as amplitude fluctuations in the function generator and the RF switches employed. A calculation of the fundamental photon shot noise-limited noise density for this system yields 0.4 $\upmu$rad$/\sqrt{\mathrm{Hz}}$ for a 18 mm$^3$ vapor cell, highlighting the potential for further improvements.

Single-optical-axis magnetometers are generally vulnerable to deadzones, depending on the orientation of the magnetic field relative to this axis \cite{bloom1962principles}. To evaluate the angular dependence of the noise, we have binned the noise density, $S_{\theta}$ into bins of polar angle, $\beta$ $\sim$ 12$^{\circ}$ each and calculated the dependence of noise density over the polar angle (Fig.~\ref{fig:data}d). Error bars indicate 1$\sigma$  variation of noise density within each bin. Across all polar angles, $S_\theta$ exhibits a modest dependence on $\beta$ and the mean noise density remains between 19 - 27 $\upmu$rad$/\sqrt{\mathrm{Hz}}$.  Notably, no divergence in noise is observed for any orientation, demonstrating deadzone-free operation. Combining measurements across six PEs compensates for orientations where individual PEs exhibit reduced sensitivity, thereby enhancing overall angular precision. Extending this approach to a larger set of PEs with differing relative phases among the three coil axes offers a promising route to further reduce angular variation of noise density and improve precision.

\section{Conclusion}

In this work, we have realized a vector OPM with a microfabricated $^{87}$Rb vapor cell capable of accurate, deadzone-free measurements of the magnetic field direction with a single optical axis. The approach combines a calibration procedure based on controlled DC magnetic field rotations with a Floquet-based theoretical model. Our proof-of-principle results demonstrate a mean angular accuracy of 80 $\upmu$rad and a mean angular noise density of 22 $\upmu$rad$/\sqrt{\mathrm{Hz}}$ over the full solid angle.  Current performance is limited in part by residual systematic effects not fully accounted for by our model. These residual direction-dependent systematics, reaching up to a few Hz and contributing  $\sim$ 30 $\upmu$rad systematic errors in vector measurement, include shifts arising from $\sigma^-$ contributions associated with residual rubidium population in $F=1$, and distortions of the Rabi lineshape due to the dynamic heading error that shift the Rabi frequency beyond the assumptions of weighted Rabi-amplitude averaging (Appendix \ref{sec:exact}). The former contribution is expected to be comparatively smaller due to the high $F=2$ spin polarization produced by synchronous optical pumping, and the smaller spin projection of the $F=1$ manifold in the Faraday rotation signal. A more complete treatment of the full Rabi lineshape therefore offers the most promising route towards reducing the remaining systematic error. Furthermore, slow technical drifts, particularly over $\mathbf{C}_t$, degrade PE calibration, contributing $\sim$ 30 $\upmu$rad errors in vector measurements, as quantified by the delay-dependent Rabi frequency drift measurements in Appendix \ref{sec:Drift}. Mitigation of these effects through enhanced RF stability and  higher bandwidth current controller, which would allow $\mathbf{C}_t$ to be reduced from $\sim$ 30 s to 9 s offers a pathway towards further improvements in performance. An additional uncertainty of $\sim$ 20 $\upmu$rad arises from the scalar calibration of the coil system used to benchmark the magnetometer (Appendix \ref{sec:scalar calibration}).

In comparison, other coil modulation techniques with a single optical axis have demonstrated excellent angular accuracies ($\sim10\,\upmu$rad) \cite{gravrand2001calibration,leger2009swarm,rutkowski2014towards} with magnetic field direction inferred directly from Larmor modulation amplitudes. In that approach, the accuracy evaluation was based on calibration of the magnetometer with sensor rotations within a static, uniform, external geomagnetic field.  Such calibration takes an extended duration, with $\mathbf{C}_t\sim$ 3 hours reported in Ref.~\cite{gravrand2001calibration} and requires the external field to be stable throughout $\mathbf{C}_t$. In our work, magnetometer calibration is instead based on external magnetic field rotations with $\mathbf{C}_t\sim$ 30 s.

Implementing  field rotation-based calibration for the coil modulation techniques with low-frequency modulations is challenging, as fluctuations within the modulation bandwidth from the coil-generated fields are amplified onto the modulation amplitudes. Fluctuations in the coil fields are effectively scaled by a factor, $B/a$ onto the measured modulation, where $B$ and $a$ denote the 
magnetic field strength ($\sim$ 50 $\upmu$T) and the modulation amplitude ($\sim$ 50 nT), respectively \cite{gravrand2001calibration}, leading to errors in the inferred direction. Our frequency-based readout, combined with our faster calibration procedure and theoretical model, enables accurate vector measurements without physical rotation of the sensor during operation. As a result, our approach reduces mechanical complexity and relaxes stability requirements on the external coil system, although residual systematics fundamentally limit the achievable angular accuracy in $^{87}$Rb relative to state-of-the-art coil modulation methods. 

While the present apparatus is not optimized for size, weight, or power, the sensing element is already compatible with miniaturized implementations through the use of a microfabricated vapor cell. Several additional modifications could further reduce the size and power requirements of the full system. The $\approx$ 400 mW pump power used here is a peak power during synchronous optical pumping, not a continuous optical power requirement. By concentrating the optical power into synchronous pump pulses with 8$\%$ duty cycle, the average pump power requirement can be substantially reduced \cite{agnesi2012low}. Furthermore, the 2 W optical heating used in this proof-of-principle setup could be reduced through improved thermal isolation \cite{li2023laser}, direct optical heating of the vapor cell \cite{mhaskar2012low}, or magnetically quiet resistive heating designed to avoid perturbing the Rabi dynamics \cite{schwindt2004chip}. Additionally, the RF signal chain could be integrated using compact RF oscillators and RF switches \cite{kitching2018chip}, while the RF coils could be implemented with printed circuit or other planar coil geometries \cite{tayler2022miniature}.

In this work, we have exclusively employed $\Omega_{\sigma+}$ measurements to determine the magnetic field direction. A natural extension is to also incorporate $\Omega_{\sigma-}$ Rabi frequencies from the  $F = 1$ hyperfine manifold, which  would provide further independent angular information on the magnetic field. In a hybrid microwave-RF configuration, atomic population pumped to $F = 2$ can be partially transferred to $F = 1$ through microwave adiabatic rapid passage, with microwaves tuned close to the ground state hyperfine splitting. This approach would allow simultaneous $\Omega_{\sigma\pm}$ measurements without increasing overall measurement time. Recent work has shown that $\Omega_\pi$ Rabi frequencies associated with the RF field can also be measured via two-photon RF transitions \cite{maddox2023two,geng2021laser}. Combining $\Omega_{\sigma\pm,\pi}$ measurements would add redundant angular information that would improve robustness against angular regions with poor sensitivity and provide internal consistency checks on the inferred direction. Such redundancy could also support drift monitoring and trigger recalibration once the drift exceeds a specified threshold \cite{kiehl2025accurate}. $\Omega_{\sigma\pm,\pi}$ measurements would also enable self-calibration based on PE structure itself \cite{thiele2018self}, using only atomic signals without external magnetic field or sensor rotations, advancing towards an electromagnetically referenced RF-based vector magnetometer that does not rely on an external coil system reference. 

We believe that the resonant Zeeman Rabi approach introduced here can be improved by extending the technique to other atomic species with more favorable level structure. In particular, metastable $^4$He, which is used in the coil modulation methods described earlier \cite{gravrand2001calibration}, has zero nuclear spin and therefore lacks hyperfine structure. As a result of selection rules, Zeeman Rabi oscillations would be driven only by the  $\sigma^+$ component of the PE, preventing residual $\sigma^-$ effects, and absence of the nonlinear Zeeman effect would prevent dynamic heading error systematics \cite{guttin1994isotropic}. Conversely, one may deliberately exploit enhanced nonlinear Zeeman effects to excite selective transitions. In particular, $^{39}$K atoms exhibit a substantially larger nonlinear Zeeman splitting between adjacent Zeeman resonances compared to rubidium and cesium at the same magnetic field. The resulting splitting can be sufficient at geomagnetic fields to resolve individual Zeeman resonances and enable selective excitation of a single $\Delta m_F = 1$ transition, reducing residual systematic effects \cite{ding2022active}. 

Beyond DC vector magnetometry, our approach can be generalized to RF metrology. The calibration protocol developed here, when combined with our Floquet-based mapping of magnetic field orientations to PE structure, can be used to infer amplitude, polarization and phase of an unknown RF field \cite{motamedi2023magnetic,kitching2025atom}. Such capability offers potential improvements for numerous applications using RF-based OPMs, including magnetic induction imaging \cite{deans2021electromagnetic,maddox2023two}, tomography \cite{marmugi2016optical}, and  ultralow-frequency RF communication \cite{fan2022magnetic,gerginov2017prospects}. 

\section*{Acknowledgements}
We acknowledge helpful discussions with Ricardo Jim\'enez-Mart\'inez, Georg Bison, and Tobias Thiele, and technical expertise from Terry Brown, Ivan Ryger, and Felix Vietmeyer. This work was supported by NSF QLCI Award No. OMA-2016244, DARPA through Grant No. W911NF-21-1-0127, and the Baur-SPIE Chair in Optical Physics and Photonics at JILA.

\section*{Data availability}

The data that support the findings of this article are openly available \cite{menon_2026_20122775}.

\appendix
\renewcommand{\appendixname}{APPENDIX}

\renewcommand{\arraystretch}{1.2}
\begin{table*}[tbh]
\caption{Summary of symbols used in this work}
\label{tab:symbols}
\begin{ruledtabular}
\begin{tabular}{c l}
Symbol & Description \\ \hline
$\mathbf{B}_{\mathrm{DC}}$ & DC magnetic field \\
$\Bcal_\mathrm{RF}\mathrm{(t)}$ & RF magnetic field \\
$\Bcal_{0}$ & Complex amplitude vector, $\sum\left(\mathcal{B}_0\right)_j\,e^{-i\phi_{j}}\,\hat{j},\;j\in \{x,y,z\}$ associated with $\Bcal_\mathrm{RF}\mathrm{(t)}$\\
$\nu_L$ & Larmor frequency \\
$\Omega_{\sigma+}$ & Rabi frequency driven by $\sigma^+$ component of $\Bcal_\mathrm{RF}\mathrm{(t)}$\\
$\gamma$ & Gyromagnetic ratio \\
$\alpha,\,\beta$ & Azimuthal and polar angles specifying DC field orientation \\
$\mu_{\sigma\pm}$ & \parbox[t]{0.8\textwidth}{\raggedright Magnetic transition dipole moment describing Zeeman Rabi oscillations within $F=2$ ($\sigma^+$) and $F=1$ ($\sigma^-$) manifolds, coupling to $\sigma^+$ and $\sigma^-$ components of the RF field, respectively}\\
$H(t)$ & Time-dependent Hamiltonian governing Rabi dynamics\\
$H_F(t)$ & Projection of $H(t)$ onto hyperfine manifold, $F$\\
$q_{F,m_F;n},\,\ket{q_{F,m_F;n}}$ & \parbox[t]{0.8\textwidth}{\raggedright Floquet quasienergy and corresponding eigenstate associated with  $|F,m_F\rangle$ and harmonic index, $n$}\\
$\widetilde{H}_F$ & Floquet Hamiltonian corresponding to $H_F(t)$\\
$\mathbf{C}_t$ & Calibration time \\
$\delta\theta$ & Angular accuracy \\
$\updelta\mathrm{B}_\mathrm{x},\updelta\mathrm{B}_\mathrm{y}$ & Transverse component errors\\
$\mathrm{S_{B_x},S_{B_y}}$ & Transverse component noise density\\
$S_\theta$ & Angular noise density\\
$\Delta\uptheta$ & Relative angular error
\end{tabular}
\end{ruledtabular}
\end{table*}
\renewcommand{\arraystretch}{1.0}
\section{\MakeUppercase{Additional Details\\ of the Experimental setup}}

The DC and RF coil systems consist of two concentric triaxial assemblies, each consisting of three nominally-orthogonal coil pairs. The coils are wound with Litz wire on  rigid non-magnetic 3D printed mounts fabricated with glass-bead-filled Nylon 12 for mechanical stability and reduce thermal variations. The vapor cell is located at the midpoint of each coil pair, so that the applied fields are maximally uniform over the rubidium atoms. The DC coil system has an average diameter  of 81 mm  and a coil factor of 0.95 $\upmu$T/mA, while the RF system has an average diameter of 39 mm with a coil factor of 0.77 $\upmu$T/mA. The average inductances of the DC and RF coil systems are 820 $\upmu$H and 56 $\upmu$H, respectively.

To maintain compatibility with microwave Rabi-driven protocols, the microfabricated cell is enclosed within a non-metallic alumina dielectric resonator cube (12.5 mm), with $\mathrm{TE}_{111}^{(x,y,z)}$ modes tuned to the hyperfine resonance of 6.834 GHz. Optical heating of the cell is implemented  using a 1550 nm heating laser and a pair of dichroic filters (Schott RG9) bonded  to two opposing faces of alumina cube along the optical axis. The filters transmit the pump and probe laser wavelengths with $\sim90\%$ transmittance while strongly absorbing the heating laser ($\sim4\%$ transmittance). To heat the cell optically, the 2 W heating laser is split into two equal parts and is incident onto each filter from opposite sides. The cell temperature is inferred from independent measurements of the hyperfine pressure shift \cite{kiehl2024correcting,vanier1982hyperfine}. This fully optical approach \cite{mhaskar2012low}, along with the thermal mass of the alumina enclosure, provides passive temperature stabilization at 100$^{\circ}$C that avoids conductive or magnetic elements that could perturb the Rabi dynamics of the rubidium atoms.

\label{sec:setup appendix}

\section{\MakeUppercase{Calibration of DC Coil System }}
\label{sec:scalar calibration}
\begin{figure}[tbh]
    \centering
    \includegraphics[width=0.47\textwidth]{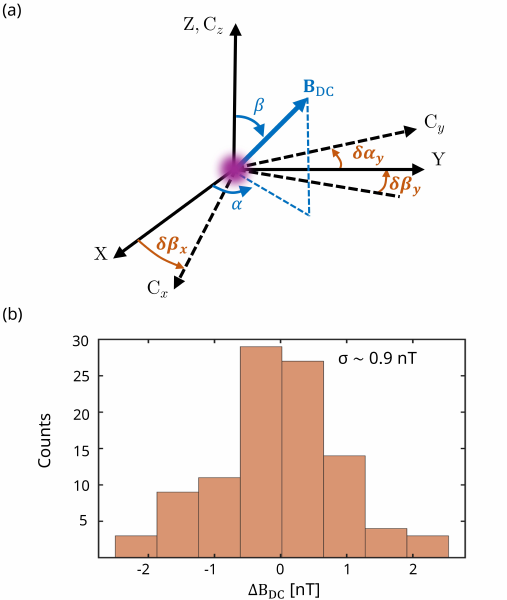}
    \caption{Scalar calibration of the DC coil system. (a) Geometry of the coil frame coordinate system C$_x$-C$_y$-C$_z$, and the orthogonalized laboratory frame coordinate system X-Y-Z. $\{\delta\beta_x$, $\delta\beta_y$, $\delta\alpha_y\}$ characterize the deviation of the coil frame from the laboratory frame. (b) Histogram of residuals resulting from calibration of the DC coil system with Larmor frequency measurements along 100 random magnetic field directions.}

    \label{fig:scalar calibration}
\end{figure}
 The DC magnetic field produced by each coil pair can be modeled as 
\begin{equation}
(\widetilde{\mathbf{B}}_{\mathrm{DC}})_j = a_jI_j\,\hat{c}_{j},\, j \in \{x,y,z\},
\end{equation}
where $a_j$ denotes the corresponding coil factor, $I_j$ is the applied current, and $\hat{c}_j$ denotes the effective axis of each coil pair (Fig.~\ref{fig:scalar calibration}a). Due to imperfections in the coil system, $\hat{c}_j$ may not be perfectly orthogonal to each other. We can express these deviations from a Cartesian coordinate system as small angular perturbations in the following manner
\begin{align}
\hat{c}_z&=  \hat{z},\notag\\
\hat{c}_x &= R_y(\pi/2 + \delta\beta_x)\hat{z},\notag\\
\hat{c}_y &=R_z(\pi/2 + \delta\alpha_y)R_y(\pi/2 + \delta\beta_y)\hat{z}.
\end{align}

Here $R_x(\theta)$ and $R_y(\theta)$ are standard rotation matrices about the $\hat{x}$ and $\hat{y}$ directions in the laboratory frame (Fig.~\ref{fig:scalar calibration}a). The small angles $\delta\beta_x,\,\delta\alpha_y,\,\delta\beta_y$ quantify the misalignment of the X and Y coils with respect to the nominal orthogonal frame, as illustrated in Fig.~\ref{fig:scalar calibration}. 

Additionally, due to residual shielding imperfections,  a background magnetic field $\mathbf{B}_{\mathrm{bg}}$ is also included in the  total magnetic field model,
\begin{equation}
\widetilde{\mathbf{B}}_{\mathrm{DC}}(I_{x},I_{y},I_{z};\{U\}) = \Sigma_j a_jI_j\,\hat{c}_{j} + \mathbf{B}_\mathrm{bg},
\label{eq:DCfield}
\end{equation}
with the full set of calibration parameters denoted by $\{U\} = \{a_x,a_y,a_z,\delta\beta_x,\delta\alpha_y,\delta\beta_y,\mathbf{B}_\mathrm{bg}\}$ 

We determine $\{U\}$ by minimizing the following cost function, $\mathcal{C}\left(\{U\}\right)$ that quantifies the error between the expected field strength based on Eq.~\eqref{eq:DCfield} and those from scalar magnetic field measurements, $\mathrm{B_{DC}}$ based on Larmor frequency measurements, $\nu_L$ for various magnetic field directions \cite{merayo2000scalar,thiele2018self,kiehl2025accurate}.
\begin{align}
    \mathcal{C}\left(\{U\}\right) = \sum_i\Bigl|\;\widetilde{\mathrm{B}}_\mathrm{DC}\Bigl(I_{x,i},&I_{y,i},I_{z,i};\{U\}\Bigl) -  \mathrm{B_{DC,i}} \notag\\& - \updelta \mathrm{B_{he}}\left(\rho_0,\beta_{nom,i}\right)\Bigr|^2,
\end{align}
where $\mathrm{B_{DC,i}}$ are the magnetic fields extracted using the procedure described in Appendix \ref{sec:Rabi frequency extraction} from the measured Larmor frequencies, $\nu_{L,i}$ for the $i^{th}$ applied magnetic field direction. A limited static heading error correction, $\updelta \mathrm{B_{he}}\left(\rho_0,\beta_{nom,i}\right)$ is applied using the initial density matrix, $\rho_0$ and at the nominal polar angle, $\beta_{nom}$, which is computed assuming an orthogonal coil system.  The resulting calibration residuals for 100 different magnetic field directions are shown in Fig.~\ref{fig:scalar calibration}b. The measured standard deviation of the residuals, $\sim$ 0.9 nT, is substantially larger than the statistical noise floor of the Larmor measurements, suggesting the presence of systematic errors unaccounted in the total field model. For 50 $\upmu$T fields, the residual level corresponds to an effective angular accuracy of approximately $0.9\,\mathrm{nT}/50\,\upmu\mathrm{T}\sim20\,\upmu$rad.

\section{\MakeUppercase{Details of the Floquet formalism}}

We show the matrix representation of the Floquet Hamiltonian , $\widetilde{H}_F$ defined in the Hilbert space $\ket{F,m_F}\otimes\ket{n}$ as described in Eq.~\eqref{eq:Flqmatrix}. 
\begin{equation}
\widetilde{H}_F =
\left(
\begin{array}{ccccc}
\scalebox{1}{$\ddots$} & & & & \\
& \scalebox{1}{$H^{(0)}_F$} - \hbar\omega_{RF}\scalebox{1}{$\mathbb{I}$}& \scalebox{1}{$H^{(1)}_F$} & \scalebox{1}{$0$}&\\
& & & & \\
& \scalebox{1}{$H^{(-1)}_F$} & \scalebox{1}{$H^{(0)}_F$}& \scalebox{1}{$H^{(1)}_F$} &\\
& & & &\\
&\scalebox{1}{$0$} & \scalebox{1}{$H^{(-1)}_F$} & \scalebox{1}{$H^{(0)}_F$} + \hbar\omega_{RF}\scalebox{1}{$\mathbb{I}$}&\\
 & & & &\scalebox{1}{$\ddots$}\\
\end{array}
\right)
\end{equation}

Here $H_F^{(n)}$ are block matrices defined in Eq.~\eqref{eq:Flqmatrix} while $ \mathbb{I}$ is the identity matrix of size $(2F+1) \times (2F+1)$. $\widetilde{H}_F$ captures both resonant couplings, $\ket{F,m_F;n}\leftrightarrow\ket{F,m_F-1;n+1}$, as well as off-resonant counter-rotating virtual couplings $\ket{F,m_F;n}\leftrightarrow\ket{F,m_F-1;n-1}$, the latter being responsible for Bloch-Siegert effects and deviations from the rotating wave approximation.

Eigenvalues of the Floquet Hamiltonian determine the Rabi frequencies via Eq.~\eqref{eq:eig}, where $\Omega_m = \delta q_m/h$ and $\delta q_m = q_{F,m;-m} - q_{F,(m-1);-(m-1)}$. For the $F = 2$ manifold, this yields four distinct Rabi frequencies $\Omega_{\sigma+,m}, m\in\{2,1,0,-1\}$ due to nonlinear Zeeman shifts.

\section{\MakeUppercase{Pumping Model}}
\label{sec:pump}
We employ a synchronous pumping scheme in which  a circularly polarized pump laser is amplitude modulated at the Larmor frequency, $\nu_L$ using a square wave modulation with a 8$\%$ duty cycle \cite{hunter2018free,hewatt2025investigating}. While the modulation frequency matches the frequency of the RF driving fields, during the pumping phase, all RF fields are turned off.

 To calculate the density matrix at the end of the optical pumping phase, $\rho_{0}$, we simulate the atomic dynamics using the following Lindblad master equation that accounts for coherent evolution and key decoherence effects \cite{kiehl2023coherence,seltzer2008developments}

 \begin{equation}
\begin{split}
   \frac{d\rho}{dt}=&\frac{\left[\,H,\rho\,\right]} {i\hbar} + \frac{\left[\delta E_{se}\langle \mathbf{S}\rangle \cdot\mathbf{S},\rho\right]}{i\hbar}\\& +\Gamma_{se}\left(\varphi_{n}\left(1+4\langle \mathbf{S}\rangle \cdot\mathbf{S}\right)-\rho\right) + \Gamma_{sd}\left(\varphi_n-\rho\right)\\&\hspace{2em}+\Gamma_{wc}\left(\rho_{e} -\rho\right) + R_{OP}\left(\varphi_{n}\left(1+2S_z\right)-\rho\right),
   \label{eq:master}
\end{split}
\end{equation}
where $H$ is the system Hamiltonian defined in Eq.~\eqref{eq:H}, with $\Bcal_\mathrm{RF}\left(\mathrm{t}\right) = 0$. Here, $\varphi_n = \rho/4+\mathbf{S}\cdot\rho\mathbf{S}$ represents the nuclear component of the density matrix \cite{seltzer2008developments}, while $\mathbf{S}$ is the electronic spin, with $\langle\varphi_n\mathbf{S}\rangle$ = 0. The rest of the terms describe spin-exchange shift ($\delta E_{se}/\hbar \sim1400\,\mathrm{s^{-1}}$ ) \cite{kiehl2024correcting,micalizio2006spin},  spin-exchange decoherence ($\Gamma_{se}\sim3890\,\mathrm{s^{-1}}$) \cite{appelt1998theory}, spin destruction due to Rb-Rb  and Rb-N$_2$ collisions ($\Gamma_{sd}\sim 80\,\mathrm{s^{-1}}$) \cite{baranga1998polarization}, spin destruction due to collisions between Rb and the inner wall of the microfabricated vapor cell ($\Gamma_{wc}\sim220\,\mathrm{s^{-1}}$) \cite{pouliot2021accurate}, and optical pumping with a perfect circularly polarized beam along the the z-axis, respectively \cite{hewatt2025investigating}. The given shift and decoherence rates are calculated for a cell size of dimensions $3\times3\times2$ mm$^3$ at a temperature of 100 $^{\circ}$C.

\section{\MakeUppercase{Effect of varying the phase offset of RF field}}
\label{sec:RF_phase}
Consider the RF field,
\begin{equation}
\Bcal_\mathrm{RF}(t) = \mathfrak{R}\left(\Bcal_0e^{-i(\omega_{RF}t+\phi)}\right).
\label{eq:RF field}
\end{equation}
Here $\Bcal_0$ specifies the complex amplitude and direction of the RF field, and $t$ = 0 defines the start of the probing sequence. $\phi$ is a phase offset applied to the RF field.

The transformation to a rotating frame can be described by the Euler-Rodrigues formula \cite{pina2011rotations}. The rotation of a vector, $\mathbf{v}$ by angle $\theta$ about an axis $\hat{b}$ is given by 
\begin{equation}
\mathbf{v}' = \mathbf{v}\cos(\theta) + (\hat{b}\times\mathbf{v})\sin(\theta) + \hat{b}(\hat{b}\cdot\mathbf{v}) (1-\cos(\theta)).
\label{eq:rodrigues}
\end{equation}

 The rotating frame is defined by rotation about an axis defined by $\hat{b} = \mathbf{B}_{\mathrm{DC}}/\mathrm{B}_{\mathrm{DC}}$ at angular frequency $\omega_{RF}$. Here we focus the $\sigma^+$ polarization component of $\Bcal_\mathrm{RF}(t)$
\begin{equation}
\left(\Bcal_\mathrm{RF}(t)\right)_{\sigma+} = \left(\Bcal_\mathrm{RF}(t)\cdot\hat{\epsilon}^*_{\sigma+}\right)\;\hat{\epsilon}_{\sigma+},
\end{equation}
where we assume both $\Bcal(t)$ and $\hat{\epsilon}_{\sigma+}$ are expressed in the laboratory frame. Transformation to a frame rotating at $\omega_{RF}$ corresponds to a rotation in the opposite sense relative to the laboratory frame, such that $\theta = -\omega_{RF}t$ in Eq.~\eqref{eq:rodrigues}. Because $\hat{b}\cdot\hat{\epsilon}^*_{\sigma+}=0$, Eq.~\eqref{eq:rodrigues} yields
\begin{multline}
\left(\Bcal'_\mathrm{RF}\right)_{\sigma+} = \left(\Bcal_\mathrm{RF}(t)\right)_{\sigma+} \cos(\omega_{RF}t) - \\\hat{b}\times\left(\Bcal_\mathrm{RF}(t)\right)_{\sigma+}\sin(\omega_{RF}t).
\end{multline}

Using $\hat{b}\times\hat{\epsilon}_{\sigma\pm} =\mp  i\hat{\epsilon}_{\sigma\pm}$ and substituting Eq.~\eqref{eq:RF field}, we obtain
\begin{align}
\left(\Bcal'_\mathrm{RF}\right)_{\sigma+}
&=\mathfrak{R}\bigg[\bigg(\Bcal_0\cdot\hat{\epsilon}_{\sigma+}\bigg) e^{-i(\omega_{RF}t+\phi)}\bigg(\hat{\epsilon}_{\sigma+}\cos(\omega_{RF}t)\notag\\&\hspace{11em}+i\hat{\epsilon}_{\sigma+}\sin(\omega_{RF}t)\bigg)\bigg]\notag\\
&=\mathfrak{R}\left[\left(\Bcal_0\cdot\hat{\epsilon}_{\sigma+}\right) \hat{\epsilon}_{\sigma+}e^{-i\phi}\right].
\end{align}

Hence, varying the phase offset by $\phi$ corresponds to a rotation of the stationary component in the rotating frame by the same angle.

\section{\MakeUppercase{Measurement of Rabi and Larmor spin dynamics}}
\label{sec:Rabi frequency extraction}
To measure Rabi and Larmor dynamics, we monitor the Faraday rotation, $\theta_f(t)$ of a far-detuned linearly polarized probe beam of wavelength 780 nm using balanced photodetection (Thorlabs PDB230A) following a polarizing beam splitter. The Faraday rotation signal provides a continuous non-destructive readout of the  spin polarization component along the probe beam propagation direction \cite{kiehl2023coherence}
\begin{equation}
\begin{split}
\theta_{f} = \frac{cr_ef_{D_2}n_{Rb}l}{2(2I+1)}&(D(\nu - \nu_{D_2,F=2})\langle F=2\rangle_Z \\&- D(\nu - \nu_{D_2,F=1})\langle F=1\rangle_Z),
\label{eq:F1}
\end{split}
\end{equation}
where $r_e$ is the electron radius, $f_{D_2}$ is the oscillator strength of the $D_2$ transition of $^{87}$Rb, $D(\nu)$ is the dispersive Lorentzian lineshape, and the propagation direction for the probe beam is along $\hat{z}$. Because the detuning of the probe beam, $\Delta\nu\sim$ 100 GHz far exceeds the hyperfine splitting, $\left|\nu_{D_2,F=2} - \nu_{D_2,F=1}\right|\sim6.834\;\mathrm{GHz}$, Eq.~\eqref{eq:F1} simplifies to 
\begin{equation}
\theta_{f}\approx\frac{cr_{e}f_{D_{2}}n_{Rb}l}{2(2I+1)}D(\nu-\nu_{D_{2}})\begin{cases}
m_{F} & F=2\\
-m_{F} & F=1
\end{cases}
\end{equation}

\begin{figure}[tbh]
    \centering
    \includegraphics[width=0.48\textwidth]{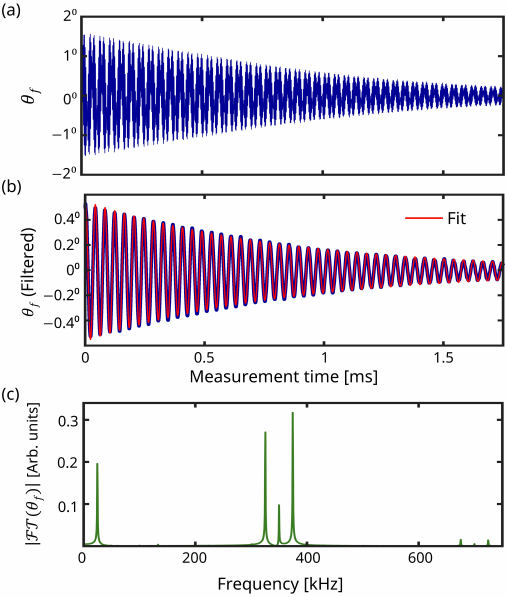}
    \caption{Extraction of Rabi frequencies from Faraday rotation measurements. (a) Raw and (b) filtered Faraday rotation measurements of Rabi oscillations driven by PE1 with the magnetic field oriented along $(\alpha,\beta)$ = $(228^\circ,71.84^\circ)$. (c) Fourier spectrum of the raw Faraday rotation signal exhibiting peaks at $\Omega_{\sigma+}$, $\nu_{RF}\pm\Omega_{\sigma+}$, and weaker features at $2\nu_{RF}\pm2\Omega_{\sigma+}$}

    \label{fig:Faraday rotation signal}
\end{figure}

The Faraday rotation signal is acquired from the balanced photodetector ports using an Alazar ATS9462 PCI Express digitizer with a sampling rate of 10 MHz. The raw Faraday rotation signal for a $\sigma^+$ Rabi oscillation with PE1, with a DC magnetic field oriented  at ($\alpha=228^\circ,\beta=71.84^\circ$) is shown in Fig.~\ref{fig:Faraday rotation signal}a. For Rabi measurements, $\theta_{f}(t)$ is filtered through a digital low pass filter with 3-dB cutoff at 170 kHz to remove components at $\nu = \mathrm{\nu_{RF}},\mathrm{\nu_{RF}}\pm\Omega_{\sigma+}$ as shown in Fig.~\ref{fig:Faraday rotation signal}b. 

The Rabi frequency, $\Omega_{\sigma+}$ is then extracted from the filtered signal by fitting the following double exponentially decaying sinusoid \cite{kiehl2023coherence,kiehl2025accurate}:
\begin{align}
\theta_f(t) = A_0 + A_{11}&e^{-t/T_{11}}+A_{12}e^{-t/T_{12}}\notag\\ &+ A_2e^{-t/T_2}\sin(2\pi\Omega_{\sigma+}t+\varphi).
\end{align}
The double exponential decay profile here reflects spin-exchange decoherence of the Faraday rotation signal \cite{kiehl2023coherence}. 

For measurement of the Larmor precession, the RF fields are turned off, and the optically pumped spin-polarized atoms are allowed to precess solely under the influence of the DC magnetic field. To extract the magnetic field, we compute the frequency spectrum using a zero-padded Fourier transform of the resulting Faraday rotation signal. This increases the density of the frequency spectrum without adding information to the underlying data. The complex Fourier spectrum  is then fit to a sum of two Lorentzian lineshapes corresponding to the $F = 1$ and $F = 2$ hyperfine manifolds. The center frequencies of the two Lorentzians are jointly constrained through the Breit-Rabi formula \cite{foot2005atomic} by a single DC field magnitude, $\mathrm{B}_\mathrm{DC}$ determined from the fit \cite{kiehl2024correcting}.

\section{\MakeUppercase{Numerical evaluation of Residual Systematics}}
\label{sec:exact}

\begin{figure}[tbh]
    \centering
    \includegraphics[width=0.48\textwidth]{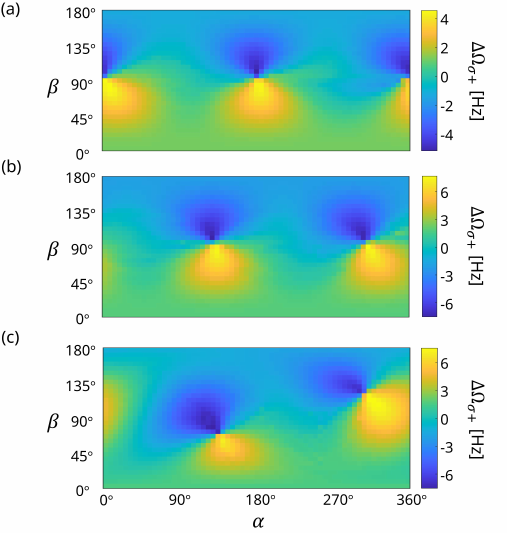}
    \caption{$\Delta\Omega_{\sigma+}$ captures the difference between Rabi frequencies simulated via the exact Hamiltonian, $H(t)$ with the Lindblad master equation, including various decoherence mechanisms such as spin-exchange and wall-collisions, and those predicted by our time-independent model (Eqs.~\eqref{eq:eig} and \eqref{eq:dynamic_he}) for (a) PE1, (b) PE4 and (c) PE6.}

    \label{fig:error}
\end{figure}

We can simulate Faraday rotation signals of Rabi oscillations by numerically solving the exact time dependent $H(t)$ given in Eq.~\eqref{eq:H}. This exact treatment avoids approximations and naturally includes Bloch-Siegert shifts, RF Stark shifts as well as systematics due to the nonlinear Zeeman effect in the Faraday rotation signal. We also include the effects of spin-exchange collisions, wall collisions and spin-destruction collisions through the Lindblad master equation given in Eq.~\eqref{eq:master}, while omitting the pumping term. Rabi frequencies are then extracted from the simulated Faraday rotation signals with the algorithm given in Appendix \ref{sec:Rabi frequency extraction}.
Comparison of Rabi frequencies extracted from the numerical simulations with those derived from our time-independent model (Eqs.~\eqref{eq:eig} and \eqref{eq:dynamic_he}) quantifies the residual systematics unaccounted for in our theoretical model. This does not include other technical systematics such as drifts that have also been observed in our experiment. Fig.~\ref{fig:error} shows the difference, $\Delta\Omega_{\sigma+}$ between the Rabi frequencies from  numerical simulations and those predicted by the time-independent model across the full solid angle for PE1, PE4, and PE6. Their effects on the average accuracy can be estimated from evaluating
\begin{equation}
\delta\theta\approx\sqrt{(\sin(\beta)\delta\alpha)^2 + (\delta\beta)^2},
\end{equation}
where $(\delta\alpha,\delta\beta) = \sum_i(\Delta\Omega_{\sigma+})_i/(\partial\Omega_{\sigma+}/\partial(\alpha,\beta))_i$, and $i$ denotes the PE index. Averaging over the solid angle yields an estimated contribution of approximately $\langle\delta\theta\rangle\sim$ 30 $\upmu$rad from these residual systematics.

One contribution to this discrepancy arises from residual population in $F = 1$ resulting from imperfect optical pumping. This population couples to the $\sigma^-$ component of the PE, producing a weak Rabi oscillation between adjacent Zeeman sublevels in $F=1$ manifold with $\Omega_{\sigma-}$ frequency that is not accounted for in the Rabi frequency extraction procedure (Appendix \ref{sec:Rabi frequency extraction}). The impact of this effect, however, is suppressed as the Faraday rotation signal from $F = 1$ is reduced by a factor of two compared to $F = 2$, reflecting smaller spin projection, $\langle F=1\rangle_z$. Additional contributions to residual systematic error lies in distortion of the Rabi lineshape due to the dynamic heading error. Analogous to the static case, Rabi dynamics with $F=2$ manifold consist of four distinct frequency components, $\Omega_{\sigma+,m},\,m\in\{2,1,0,-1\}$ in Eq.~\eqref{eq:eig},   corresponding to four overlapping Lorentzian peaks with their amplitudes determined by $A_m$ in Eq.~\eqref{eq:Amplitudes}. Their superposition can distort the observed Rabi lineshape such that the extracted single Rabi frequency is not fully described by the Rabi amplitude-weighted averaging. This is particularly true as the decoherence rate, $\Gamma$ increases and the individual Lorentzians broaden.

\begin{figure}[tbh]
    \centering
    \includegraphics[width=0.46\textwidth]{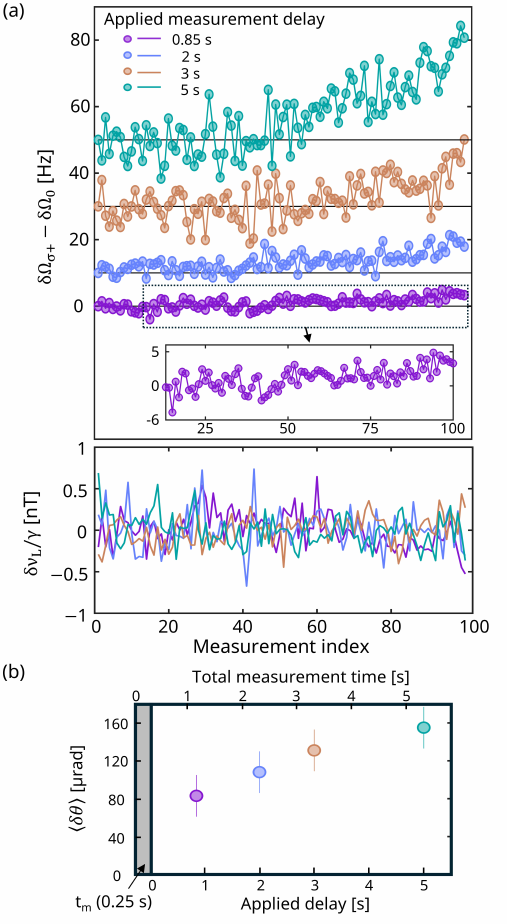}
    \caption{Effect of measurement delay on Rabi frequency drift and  vector performance. (a) Rabi frequency drift for PE6, $\delta\Omega_{\sigma+}[n] = \Omega_{\sigma+}[n] - \Omega_{\sigma+}[1]$ where $n$ is the measurement index. $n$ denotes the $n^{th}$ consecutive Rabi frequency measurement spaced by the applied delay. Inset shows zoomed view of the Rabi frequency drift at the standard delay of 0.85 s. Bottom plot shows the corresponding Larmor frequency drift, $\delta\nu_L[n]/\gamma = (\nu_L[n]-\nu_L[1])/\gamma$.  For clarity, the Rabi frequency time series are vertically offset by an arbitrary reference $\delta\Omega_0$, indicated by horizontal black lines. The average Rabi frequency, $\langle\Omega_{\sigma+}\rangle$ is 56.089 kHz across the 100  measurements. (b) Average angular accuracy, $\langle\delta\theta\rangle$ measured for over 340 applied field directions as a function of applied delay.}

    \label{fig:drift}
\end{figure}

\section{\MakeUppercase{Dependence of RF Drifts on Applied Measurement Delay}}
\label{sec:Drift}
During both the calibration and vector measurement sequences, we apply a delay of 0.85 s at the start of the measurement (Fig.~\ref{fig:timing}). This delay ensures that the DC current and  correspondingly the magnetic field at the cell has reached steady state before the beginning of the measurement. This delay however increases the total duration of each sequence, and allows drifts to accumulate across the calibration time. In our setup, the dominant drifts manifest as slow variations in the RF amplitudes.  

We quantify the influence of these drifts on vector performance by varying the applied delay and measuring both the Rabi frequency drift and the resulting angular accuracy. We apply delays of 850 ms, 2 s, 3 s, and 5 s, and measure the Rabi frequency repeatedly for PE6 at a fixed magnetic field direction, $(\alpha, \beta) = (45^{\circ},45^{\circ})$ (Fig.~\ref{fig:drift}a). The time between successive measurements is  the sum of the applied delay and measurement time, $t_m$ of 250 ms. PE6 is chosen as it is realized using all three RF channels and thus is sensitive to drifts  in all three channel amplitudes and phases. Over 100 consecutive measurements, the drift in the Rabi frequency is observed to increase monotonically with the applied delay, although the dependence is not linear. Additionally, longer applied delays is observed to lead to larger shot-to-shot fluctuations in Rabi measurements.

By contrast, the Larmor frequency remains comparatively stable over the same time scale, with variations corresponding to $\lesssim$ 1 nT, as shown in the bottom plot of the Fig.~\ref{fig:drift}a. This indicates that the dominant drift does not originate from fluctuations of the applied DC field. Furthermore, the DC field is generated by a separate, larger Helmholtz coil system with higher current requirements, yet it exhibits better stability compared to the RF system. Therefore, the primary origin of the drifts is likely the thermal sensitivity of the components in the RF delivery chain, such as the RF switches or drive electronics, rather than heating of the RF coils. We also measure the angular accuracy for the same 342 random directions as in Sec.~\ref{sec:eval} for the same applied delays and observe a clear improvement as delay is reduced (Fig.~\ref{fig:drift}b).  Extrapolating the delay to zero yields a mean angular accuracy of $\sim50\,\upmu$rad. Compared with the measured 80 $\upmu$rad for 850 ms delay, this indicates an estimated drift-dependent contribution of $\sim30\,\upmu$rad.

Table \ref{tab:accuracybudget} summarizes the dominant sources of error that limit the magnetometer accuracy.
\begin{table}[hbt]
\caption{Angular accuracy budget}
\label{tab:accuracy}
\begin{ruledtabular}
\begin{tabular}{cc}
Source of error & Estimated contribution \\
\hline
\\[-1em]
Technical drifts (Appendix \ref{sec:Drift})& $\sim$ 30 $\upmu$rad\\
Residual systematics (Appendix \ref{sec:exact}) & $\sim$ 30 $\upmu$rad\\
\shortstack{Scalar calibration of \\ DC coil system (Appendix \ref{sec:scalar calibration})} & $\sim$ 20 $\upmu$rad\\
\end{tabular}
\end{ruledtabular}
\label{tab:accuracybudget}
\end{table}

\section{\MakeUppercase{Dependence of angular accuracy on RF amplitude}}

\begin{figure}[htb]
    \centering
    \includegraphics[width=0.48\textwidth]{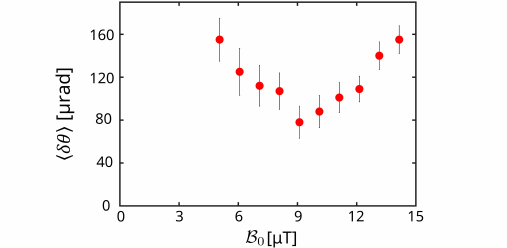}
    \caption{Variation of mean angular accuracy with amplitude of the RF field. Because of unequal coil factors, the RF field amplitudes are $\mathcal{B}_0$, $1.25\,\mathcal{B}_0$ and $0.77\,\mathcal{B}_0$ for the X, Y and Z RF coil axes, respectively.}
    \label{fig:RF amplitude}
\end{figure}

 We also study how angular accuracy depends on the RF field amplitude by measuring accuracy for a 100 random DC magnetic field directions (Fig.~\ref{fig:RF amplitude}). The measured trend reflects competing effects. The vector magnetometer is observed to be less accurate at larger RF amplitudes, likely because the increased RF power results in larger thermal drifts in the RF system, particularly in the associated RF electronics.
 Additionally, the RF current waveform becomes increasingly distorted  as the controller approaches its maximum voltage compliance while driving 350 kHz currents through an inductive load. At lower RF amplitudes, the reduced Rabi frequencies make the measurement more sensitive to residual systematics and frequency fitting errors, increasing their fractional effects and leading to a worse overall angular accuracy. As a result, we choose an amplitude of $\mathcal{B}_0 = 9.1 \,\upmu$T for our measurements. Because of differences in coil factors, the RF amplitudes for fields applied along the X, Y and Z RF coil axes are $\mathcal{B}_0$, $1.25\,\mathcal{B}_0$, and $0.77\,\mathcal{B}_0$, respectively.

\FloatBarrier
\bibliography{references}
\end{document}